\documentclass{aa}

\usepackage{graphicx}
\usepackage{txfonts}
\usepackage{makecell}

\newcommand{\equ}[1]{Eq.~\ref{eq:#1}}

\newcommand{\fig}[1]{Fig.~\ref{fig:#1}}

\newcommand{\tab}[1]{Table~\ref{tab:#1}}

\newcommand{\sect}[1]{Sect.~\ref{sec:#1}}
\newcommand{\app}[1]{Appendix~\ref{app:#1}}

\DeclareMathOperator{\sech}{sech}
\newcommand{\tig}{\ensuremath{T_{\rm 1G}}}
\newcommand{\tiig}{\ensuremath{T_{\rm 2G}}}

\begin{document}

 \title{Origin of the spectacular tidal shells of galaxy NGC\,474}
  \author{Michal B\'{i}lek \inst{1,2,3,4} \and J\'er\'emy Fensch\inst{5} \and Ivana Ebrov\'a\inst{1,6}  \and Srikanth T. Nagesh\inst{7} \and Benoit Famaey\inst{2}  \and Pierre-Alain Duc\inst{2} \and Pavel Kroupa\inst{8,9}
          }

  \institute{
        Nicolaus Copernicus Astronomical Center, Polish Academy of Sciences, Bartycka 18, 00-716 Warsaw, Poland\\
             \email{michal.bilek@asu.cas.cz}
        \and 
        Universit\'e de Strasbourg, CNRS, Observatoire astronomique de Strasbourg (ObAS), UMR 7550, 67000 Strasbourg, France
        \and
        LERMA, Observatoire de Paris, CNRS, PSL Univ., Sorbonne Univ., 75014 Paris, France
        \and
        Coll\`ege de France, 11 place Marcelin Berthelot, 75005 Paris, France
        \and
        Univ. Lyon, ENS de Lyon, Univ. Lyon 1, CNRS, Centre de Recherche Astrophysique de Lyon, UMR5574, F-69007 Lyon, France
        \and 
        FZU – Institute of Physics of the Czech Academy of Sciences, Na Slovance 1999/2, Prague 182 21, Czech Republic
        \and
        Argelander-Institut f\"ur Astronomie, Universit\"at Bonn, Auf dem H\"ugel 71, 53121 
        Bonn, Germany.
        \and
        Helmholtz-Institut f\"ur Strahlen-und Kernphysik, Universit\"at Bonn, Nussallee 14-16, D-53115 Bonn, Germany 
        \and
        Astronomical Institute, Faculty of Mathematics and Physics, Charles University in Prague, V Hole{\v s}ovi{\v c}k{\' a}ch 2, CZ-18000 Praha, Czech Republic
            }

   \date{Received ; accepted }

 
  \abstract
   {The lenticular galaxy NGC\,474 hosts a rich system of tidal shells and streams, some of which are exceptionally bright. Two teams recently presented spectroscopic observations of the brightest shells. These were the first shell{ spectra  ever observed in integrated starlight. The authors} studied the stellar populations of the shell, of the center of the galaxy, and of its globular clusters. The precise formation scenario for the tidal features of this prominent galaxy still remained unclear, however. 
   }
   {Here, we add further clues on their formation from the radii of the shells, and we present a scenario for the formation of the tidal features that seems to be unique and can explain all available data.
   }
   {Shell radii were analyzed with the shell identification method, and we ran self-consistent simulations of the formation of the tidal features. { We considered Newtonian as well as MOND gravity.}
   }
   {Observations suggest that the tidal features originate from the accretion of a spiral galaxy.
   { According to the shell identification method,} the merging galaxies first collided 1.3\,Gyr ago and then again 0.9\,Gyr ago, thereby forming the shells in two generations. This would also explain the young ages of stellar populations in the center of the galaxy and the young age of the globular clusters. The analytic models of shell propagation that underlie the shell identification method are verified by a simulation.{ The simulations reproduce the observed morphology of the tidal features well.} The accreted spiral likely reached NGC\,474 on the plane of the sky nearly radially from the south, its rotation axis pointing toward us. It probably had a stellar mass of about one-sixth of NGC\,474, that is, $10^{9.8}\,M_\sun$.    Apparently, all tidal features in the galaxy originate from one merger.}
   {}
  
   \keywords{Galaxies: individual: NGC\,474;
   Galaxies: formation; 
   Galaxies: interactions; 
   Galaxies: peculiar; 
   Methods: analytical;
   Methods: numerical}
               
   \maketitle
%

\section{Introduction}
Stellar shells in galaxies appear as arc-like glowing features with sharp outer edges. 
They are observed mainly in elliptical and lenticular galaxies, 10--20\%  of which have shells \citep{MC83,atkinson13,bil20}. 
{
They are the most common interaction signature in a complete sample of nearby massive elliptical galaxies \citep{tal09}.} 
A~few percent of spiral galaxies also have shells{
\citep{ss88,md10}. 
Currently, several hundred shell galaxies are known \citep[][and others]{MC83,thr89,ft92,forbes94,reduzzi96,tal09,ra11,atkinson13,duc15, kadofong18,bil20} including galaxies hosting quasars \citep{canalizo07,hus10,ra11}, a Local Group dwarf spheroidal \citep{coleman04}, M\,31, which is the nearest major galaxy to the Milky Way \citep{fardal07,fardal08,fardal12}, and even the Milky Way itself probably possesses multiple shells \citep{helmi03,deason13,donlon20}.
} 

There can be just one shell in a galaxy, but also up to a few dozen
{
\citep{prieur88,bil16,mil17,bil20}.} 
Shells can be aligned with the major axis of the host galaxy or be distributed randomly in azimuth \citep{wilkinson87,prieur90}. They are often accompanied by streams and tails.
Shells typically have a very low surface brightness. 
They are detected down to the current observational surface brightness limit of almost 30\,mag\,arcsec$^{-2}$ \citep[e.g.,][]{bil20}. The galaxy NGC\,474 that we study here contains some of the brightest known shells that reach about 25\,mag\,arcsec$^{-2}$ in the $R$ band \citep{turnbull99}. The relative faintness of shells {has} prevented their spectroscopic observations{ in integrated starlight}  until recently \citep{alabi20,fensch20}.{
Several measurements of shell kinematics use individual kinematic tracers, however: red giant branch stars of the so-called Western Shelf in the M31 Andromeda galaxy \citep{fardal12}; globular clusters \citep{romanowsky12} and planetary nebulae \citep{lon15b} in the giant elliptical galaxy M\,87; a combination of globular clusters, planetary nebulae, and H\,II regions in the Umbrella Galaxy NGC\,4651 \citep{foster14}; and RR Lyrae and blue horizontal branch stars in the Milky Way \citep{donlon20}. 
However, NGC\,474 is the only shell galaxy so far for which information about the shell kinematics and stellar population extracted from the spectroscopy of the integrated stellar light is available.
} 

Several scenarios of the origin of shells have been proposed; see \citet{ebrovadiz} or \citet{bilcjp} for a review. The most successful scenario, which we assume hereafter, is the phase-space wrapping model by \citet{quinn83, quinn84}. 
The formation of shells begins by a nearly radial minor or intermediate merger of two galaxies 
(more eccentric orbits tend to produce streams or tails; \citealp{amorisco15,hendel15}).
The more massive galaxy is called the primary, and the less massive galaxy is the secondary. The secondary is disrupted by tidal forces and loses most of its material when it reaches the pericenter of its orbit. The released stars in this moment have various radial velocities because of the internal velocity dispersion of the secondary. The released stars start to oscillate in the potential well of the primary, with orbital periods determined by their pericentric velocities. The shells are density waves made of stars near the apocenters of their orbits.{ The shells expand in time, and  their number in the host increases (even after accreting one secondary, and even if the secondary is completely disrupted after the first pericentric approach).}
{
The measurement of the number and radii of the shells in a galaxy can lead to estimates of the mass distribution of the host galaxy and the time since the merger \citep{quinn84,dc86,hq88,hq87a,hq87b,canalizo07,sh13}.
More information can be gained using the measurements of the shell kinematics \citep{mk98,ebrova12,sh13,bil15,dp21}. 
} 

Every shell can be assigned a serial number according to the number of pericentric passages that the stars constituting the shell have completed since their release form the secondary (the pericentric passage that caused their release from the secondary does not count). In other words, the $N$-th shell is made of stars near the $N+1$st apocenter of their orbit since the release of the stars from the secondary. If the core of the secondary survives,  it can give rise to further shell generations during the subsequent pericentric passages \citep{dc87,segdup96}. Each shell can then be assigned serial and generation numbers. Based on the image of a shell galaxy, it is not apparent what these numbers are. The shell identification method \citep{bil13} aims to recover them.

The shell identification method allows exploiting the observed shell radii to investigate the gravitational potential of a shell galaxy far from its center and to estimate the time since the shell-forming merger \citep{bil13,bil14,bilcjp,ebrova20}. The method relies on analytic models of the evolution of shell radii in a given gravitational potential. Such models were developed assuming that the shells are formed in exactly radial minor mergers.  It has long be thought that such mergers are nearly necessary for the formation of shells \citep[e.g.,][]{bildiz}.  Recent observational and theoretical results {suggest}, however, that shells often form even in relatively equal-mass mergers \citep{kadofong18,pop18}.
Moreover, simulations have shown shell formation also for nonradial mergers{ \citep{hq88, ebrova20}}. This requires verifying the analytic models of the shell propagation. We are aware of only one published self-consistent simulation that has been compared to analytic models. \citet{ebrova20} found that the analytic models work well for a nonradial merger for a large shell, but less so for small shells. Further simulations are needed to ascertain the precision or applicability of the models of shell evolution, but also to explore which other types of tidal features form together with shells, the location at which new stars form if some of the merging galaxies contain gas, and so on.

Our work has been inspired by the recent publication of the first  shell spectra ever{ observed in integrated starlight}.  Two teams independently published  {integrated-field-unit} spectra of one of the shells of the galaxy NGC\,474.  \citet{alabi20} used {the Cosmic Web Imager Integral Field Spectrograph mounted at Keck} and \citet{fensch20} used {the 
Multi Unit Spectroscopic Explorer mounted at the Very Large Telescope}. The galaxy hosts many shells and streams, some of which are exceptionally bright. They were the reason that NGC\,474 was listed in the first atlas of peculiar galaxies \citep{arp66}. \citet{alabi20} and \citet{fensch20} studied the star formation histories and chemical properties of the shell, of the associated globular clusters, and of the center of the galaxy. \citet{fensch20} also studied the kinematics of the shell and of the associated globular clusters and planetary nebulae. It remained unclear at which time the shells in NGC\,474 formed, whether all tidal features in the galaxy formed together, and whether the tidal features have any relation to the young stellar populations found in the galaxy.

The aim of this paper is to further our knowledge of the formation of this prototypical shell galaxy. We apply the shell identification method to estimate  the time since shell formation. We combine this with other observational characteristics of the galaxy.   We also perform self-consistent $N$-body simulations that reproduce the morphological features seen in NGC\,474.{ We consider both Newtonian and MOND \citep{milg83a} gravity models.} The result seems to be a relatively unambiguous reconstruction of the late history of the galaxy that explains the formation of the tidal features and of the young stellar populations. Using simulations, we also verify that the analytic models of shell radii evolution, underlying the shell identification method, work well for shell systems that look like the system in NGC\,474.

The paper is organized as follows. In \sect{obs} we summarize the observational characteristics of NGC\,474 and of the group in which it resides. We find many pieces of evidence that the galaxy accreted a spiral companion.{ In \sect{methods} we describe the tools we used, that is, the analytic shell identification method, described in \sect{modeling}, and self-consistent $N$-body simulations, described in \sect{sims}. We perform the analysis both with MOND modified gravity and with Newtonian gravity with dark matter. The analysis with MOND is presented in \sect{mondpot}. We begin  in \sect{ident} by finding the shell serial and generation numbers and the time since the encounter using the shell identification method. In \sect{sim} we then perform a MOND simulation reproducing the observed morphology and kinematics of the tidal features in NGC\,474. The nature of several special shells is discussed in \sect{nature}.  The results then allow us to explain in \sect{stelpop} how the young stellar population in NGC\,474 formed. In \sect{init} {we point out that assuming the MOND theory, there are strong indications that} the merger that formed the shells in NGC\,474 was initiated by galaxy NGC\,470. The Newtonian analysis is presented in \sect{nwt}. Because the parameters of the dark matter halo of NGC\,474 are not known exactly, we apply in \sect{dmpot} the shell identification method assuming a series of several credible halos. We find that for all of them, the shell identification found for MOND is possible as well, but the there is a considerable uncertainty of the time since the merger. A Newtonian simulation reproducing much of the morphology of NGC\,474 is presented in \sect{nwtsim}. We note in \sect{nwtinit} that the role of NGC\,470 in the formation of shells in NGC\,474 is speculative when Newtonian gravity is assumed. We summarize and synthesize our results in \sect{sum}. Additional information is provided in the appendices.  Technical details of our simulations are described in \app{details}. In \app{ansim} we verify that the analytic models of the evolution of shell radii on which the shell identification method is based work well with the MOND simulation. We discuss an alternative scenario of the formation of the shells that is indicated by an analysis of the shell radii in \app{alternative}, but we find that the scenario does not agree well with the other constraints we have. Finally, in \app{other} we verify that our MOND estimates of the ages of the shells in NGC\,474 do not depend much on the uncertain parameters of the gravitational potential of the galaxy.}

\begin{figure}
        \resizebox{\hsize}{!}{\includegraphics{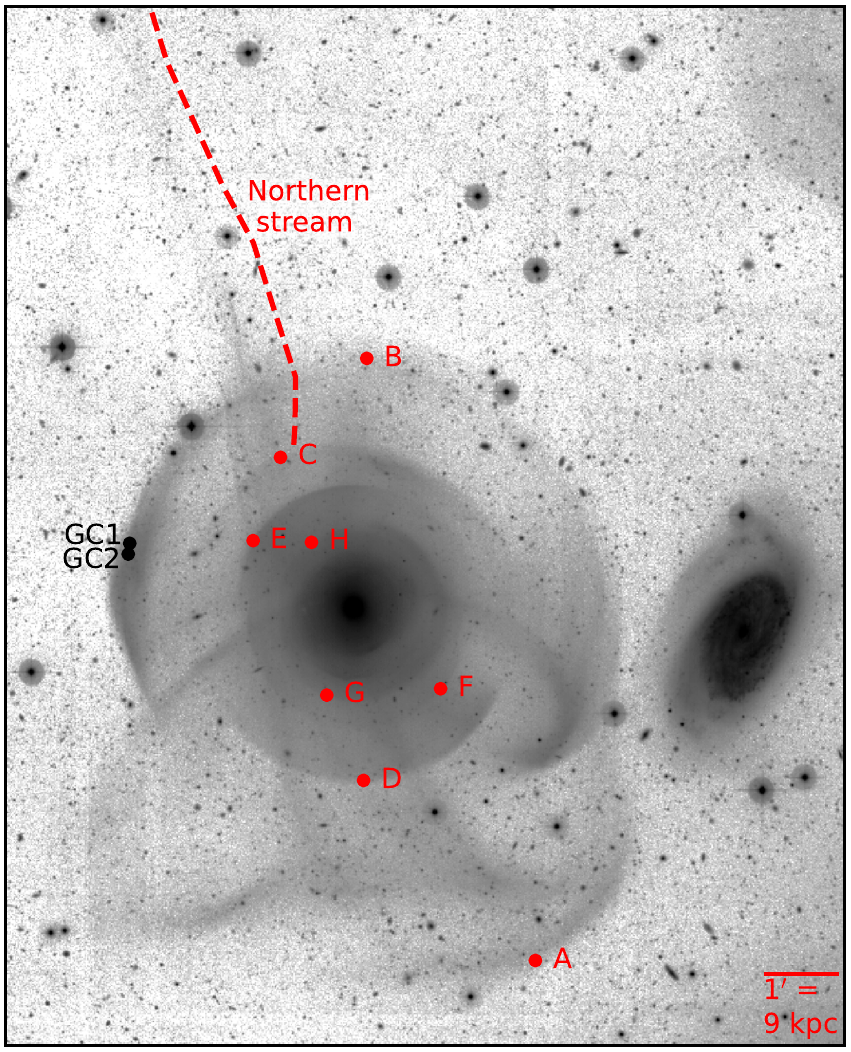}}
        \caption{Tidal features in the galaxy NGC\,474. The{ red} points indicate the positions{ used to measure the shell radii}. We also indicate the locations of the young globular clusters GC1 and GC2 observed by \citet{fensch20}. North is up, and east is to the left. The spiral galaxy is NGC\,470. The image is taken from the MATLAS survey.
        }
        \label{fig:annot}
\end{figure}

\begin{table*}
 \centering
\caption{Characteristics of the shells in NGC\,474.}
    \label{tab:shtab}
    \begin{tabular}{lccccl}
    \hline
    Label & \makecell[c]{RA \\ {[h:m:s]}} & \makecell[c]{Dec \\ {[d:m:s]}} & \makecell[c]{Radius\\ {[arcmin]}} & \makecell[c]{Radius\\ {[kpc]}} & Identification \\\hline\hline
        A & 01:19:56.473 & +03:19:59.74 & 5.56 & 50.0 & 1I \\
        B & 01:20:05.913 & +03:28:24.64 & 3.48 & 31.3 & 2I (+1II?) \\
        C & 01:20:10.747 & +03:27:01.61 & 2.32 & 20.9 & 3I (+2II?) \\
        D & 01:20:06.101 & +03:22:30.65 & 2.43 & 21.9 & 2II (+3I?) \\
        E & 01:20:12.283 & +03:25:51.76 & 1.68 & 15.1 & 3II \\
        F & 01:20:01.783 & +03:23:47.64 & 1.67 & 15.0 & 3II \\
        G & 01:20:08.161 & +03:23:42.17 & 1.29 & 11.6 & 4II \\
        H & 01:20:09.014 & +03:25:50.36 & 1.07 & 9.7 & 5II \\\hline
    \end{tabular}
\end{table*}

\section{Observational characteristics of NGC\,474}
\label{sec:obs}
We assumed a distance to NGC\,474 of 30.9\,Mpc \citep{cappellari11a}, corresponding to an angular scale of 8.99\,kpc per arcminute.  We adopted a stellar mass of $10^{10.6}\,M_\sun$  \citep{sheth10}, an effective radius of 51\arcsec , and a S\'ersic index of  8 \citep{salo15}. The galaxy is a fast rotator \citep{emsellem11} that resides in a poor galaxy group.

The galaxy is well known for hosting at least ten shells, some of which have exceptionally high surface brightness. They were studied in detail in \citet{turnbull99} and \citet{sikkema07}. In \tab{shtab} we list the radii and other characteristics of the eight largest shells of the galaxy.
We measured their radii in the image taken in the MATLAS deep-imaging survey\footnote{Jpg versions of the images are publicly available at \url{http://obas-matlas.u-strasbg.fr/}.} \citep{duc15, bil20}. The radii were measured from the brightest pixel of the galaxy (RA = 01:20:06.677, Dec = +03:24:56.26) to the points at the edges of the shells marked in  \fig{annot}. See \tab{shtab} for the exact coordinates of the measurement points.  Every shell was assigned a label serving as its name in this paper. Structure $A$ does not have the typical morphology of a shell, but we find evidence that it is a shell below. In addition to the shells, a prominent stream is contained in the galaxy that is oriented northward from the center. The stream appears to be a continuation of a linear structure that stretches from the irregular shell $A$ to the galaxy center from the south.  Shell $B$ has been studied spectrocopically in \citet{alabi20} and \citet{fensch20}. Galaxy NGC\,474 has very bright shells up to a galactocentric distance of about 30\,kpc. At larger distances, only the very faint northern stream described above extends up to at least 80\,kpc. The MATLAS image of the galaxy revealed an overdensity of globular clusters in the regions of tidal features \citep{lim17}.

There are several indications that the shells in NGC\,474 come from an accreted gas-rich spiral galaxy. \citet{dc86} reported that {a} simulation of a merger of an elliptical primary with a spiral secondary  produced a morphology of tidal structures that is very similar to the morphology seen in NGC\,474, that is, shells whose surface brightness varies in azimuth, combined with a  long radial stream. Moreover, NGC\,474 contains HI gas \citep{schiminovich97, mancillas19} and unrelaxed dust patches in the center  \citep{sikkema07}, which suggest the accretion of a gas-rich spiral. Even more evidence has been provided by recent spectroscopic studies. \citet{fensch20} studied spectra of the center of NGC\,474 and found an old stellar population typical for an early-type galaxy mixed with a young stellar population. In addition, they found two young globular clusters in the area of shell $B$.  \citet{alabi20} discovered that the stellar population of shell $B$ continued to grow until 2\,Gyr ago, which suggests an origin of this material from a star-forming galaxy as well.

The mass of the accreted galaxy was estimated by \citet{alabi20} to be greater than $10^{9.6}\,M_\sun$ from spectroscopy and about $10^{10}\,M_\sun$ from photometry of the shells. \citet{fensch20} arrived at similar mass estimates, about $10^9\,M_\sun$ from spectroscopy and $10^{10}\,M_\sun$ from photometry. The spectroscopic estimates were based on the observed correlations of chemical abundances and stellar masses of galaxies. The spectroscopic estimates are lower than the photometric estimates, likely because the accreted secondary had a metallicity gradient and the metal-poor material from the outskirts of the disk ended in the shell that was observed.
These estimates imply a mass ratio of the merger progenitors between 1:4 and 1:40. Simulations might give further constraints on the mass ratio.

NGC\,474 forms a pair with the spiral galaxy NGC\,470. With a stellar mass of $10^{10.76}\,M_\sun$ \citep{sheth10}, NGC\,470 is more massive than  NGC\,474. The galaxies lie in projection 5.46\arcmin = 49\,kpc apart and their radial velocities are different by 172\,km\,s$^{-1}$. The real separation of the two galaxies is likely substantially larger than the projected separation because shell $B$, with a radius of 31\,kpc, is circular without any convincing sign of a tidal deformation.  The distribution of HI gas around the two galaxies suggests{ that they tidally interacted in the past because they seem to lie in a common gas cloud and the gas forms structures resembling tidal arms in NGC\,470 that point to NGC\,474 } \citep{schiminovich97,rampazzo06}.

\section{Methods}
\label{sec:methods}

\subsection{Shell identification method, and analytic models of  the evolution of the shell radii}
\label{sec:modeling}

{ The primary goal of this paper is to decipher the formation history of the shells in NGC\,474. Our main tool is the shell identification method \citep{bil13}, whose core is formed by analytic models of the evolution of the shell radii with time. The speed of the shell evolution depends on the profile of the gravitational potential of the host. It does} not depend much, for example, on the velocity of the colliding galaxies. We elucidate this fact with the help of the simplest model of shell propagation by \citet{quinn84}. Let $P(r)$ denote the orbital period of a star on a radial orbit with an apocenter at a distance $r$ from the center of the host, or, in other words, twice the free-fall time from the distance $r$. Next, we assume that the stars are released from the secondary when it passes through the center of the primary, that the shell edges consist of stars in the apocenters of their orbits, and that the apocentric distances are much larger than the size of the secondary. Then, at time $t$ after the pericentric encounter of the galaxies, the $n$-th shell is located at the radius $r$ meeting the condition 
\begin{equation}
    t = P(r)(n+1/2).
\end{equation}
This formula is not exact, for example, because shells are made of stars that reach\textit{} their apocenters. More complex models exist; see the thorough review by \citet{bildiz}. Shells with high serial numbers have low contrast and cannot be observed. When a shell becomes too large, it disappears because there are too few stars with high-enough energies to make up the shell.   

We modeled the shell evolution with the analytic model of \citet{bil14}.
The simulation of \citet{ebrova20} showed that  for the largest shells, the model works well even for shells formed in highly nonradial minor mergers.  In \app{ansim} we verify with a self-consistent simulation that this analytic model works well for a shell system with an appearance similar to the shell in  NGC\,474.

In the shell identification method, the evolution of shell radii is modeled in the assumed gravitational potential and is compared with the observed shell radii. If the assumed gravitational potential is correct{ and all of the observed shells formed in one generation}, the model reproduces the observed shell radii at a certain time, which corresponds to the time that passed since the observed shells were formed. If the shells were formed in several generations, the observed{ shells can be divided into several groups, such that the shells in each group are reproduced} by the model for a certain time. These times correspond to the ages of the individual shell generations. At the end, every shell is assigned an identification, that is, a serial and generation number, and the ages of the shell generations are recovered. 

{ In practice, we searched for the shell identification in NGC\,474 in the following way. After choosing the gravitational potential, we plotted in one figure the modeled radii of the shells as functions of time  for shell serial numbers lower than 10, and with horizontal lines, we plotted the observed shell radii. We then used this plot to determine how the observed shell radii might be sorted into groups, such that each group is reproduced by the model at a certain time. In this way, we {obtained} the shell identification. It is easiest to first determine the times at which the outermost shell is reproduced and then to verify if the radii of the smaller shells are reproduced at some of these times as well. There are some additional guides to find a shell identification that were detailed in \citet{bil13}. They are  applications of Occam's razor: we should not postulate the existence of more shell generations than necessary, and we should not postulate the existence of too many shells escaping observations. This visual approach of finding the shell identification was sufficient for NGC\,474, where we analyzed only eight shells and were interested mainly in the ages of the shell generations. \citet{bil14} presented a code that can find all possible shell identifications. The code was necessary for that work because we studied a galaxy with many more shells, and it was essential not to miss any possible shell identification. The input parameters of the code are the maximum allowed number of shell generations and of the missing shells and the maximum allowed deviation of the observed and modeled shell radii. The last parameter, which should also be considered in the visual way of shell identification, still needs to be explored systematically using simulations. When we searched for shell identifications in this paper, we assumed that the analytic models are nearly precise. }

When we identified the observed shells in NGC\,474, we did not consider shells that lay closer to the center of the galaxy than shell $H$. This is because the method becomes less reliable in the central region because of the small spacing between the neighboring shells. The method becomes sensitive to errors such as in the modeling of the gravitational potential, in measurements of the observed shell radii, or in the models of shell propagation, therefore the possible nonradiality of the merger would play a larger role for the small shells.

In the models of shell propagation in NGC\,474, we assumed gravitational potentials based either on Newtonian gravity or on the MOND modified-gravity theory. In both cases, we modeled the distribution of stars in NGC\,474 with a S\'ersic sphere with the parameters described in \sect{obs}. The density of the S\'ersic sphere was evaluated using the approximation of \citet{limaneto99} and \citet{marquez00}.

When we discuss the possible shell identifications in the following subsections, we assume that the whole shell system was formed by accretion of a single galaxy, possibly in several generations. The assumption of the origin of the shells from a single accreted galaxy is not only guided by Occam's razor. { It is also supported by our simulations.} In principle, wherever we mention two or more generations, the shells might also have been formed by two or more secondaries, but this option has no support in the data,{ at least for the less degenerate MOND-based model}.

\subsection{Simulations}
\label{sec:sims}
We performed self-consistent $N$-body simulations of mergers that form tidal structures similar to those in NGC\,474. We present here two of them: one made with MOND gravity (\sect{sim}), and another in Newtonian gravity (\sect{nwtsim}). The purpose is to demonstrate that a merger of an elliptical primary and a disk secondary, for which there are observational indications (\sect{obs}), can indeed form the observed morphology of the tidal structures and, moreover, to verify the analytic models of shell propagation (this is done in \app{ansim}).
In addition, the simulation helped us to interpret some of the observed tidal structures. We did not aim to find the best possible match of the observed galaxy. That would require a much more complete exploration of the parameter space. 
The simulation did not include gas or the neighboring galaxy NGC\,470.{ The technical details of the simulations can be found in \app{details}.}

\begin{figure}
        \resizebox{\hsize}{!}{\includegraphics{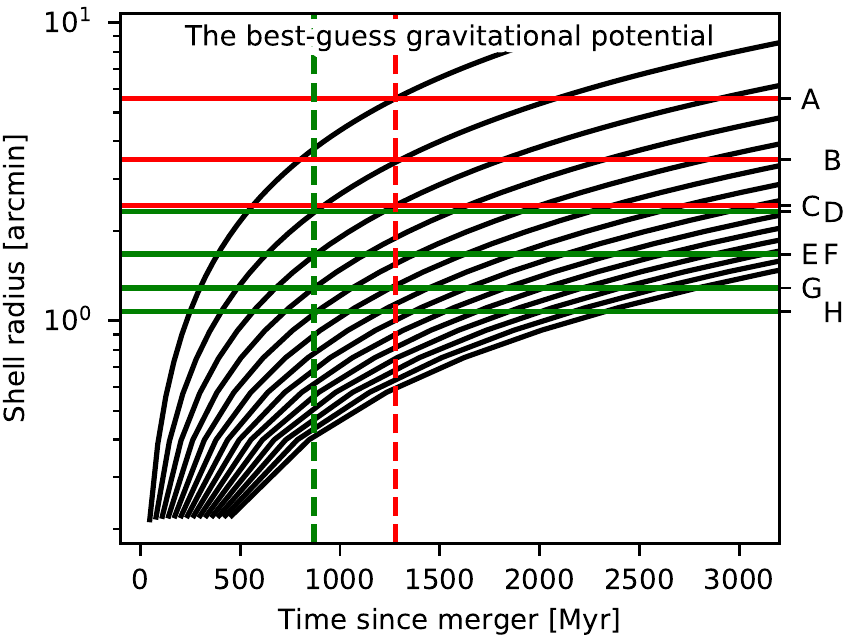}}
        \caption{{Comparison of the observed shell radii with the modelled evolution of shell radii for the gravitational potential of NGC\,474 expected by MOND. Curved lines: Modeled evolution of shell radii.} Horizontal lines: the observed shell radii. Vertical lines: Times of the best match of the model with observations, implying the times since the pericentric passages of the secondary. The two colors of the horizontal lines indicate the { adopted} separation of the observed shells into two generations.}
        \label{fig:shcomp}
\end{figure}

\section{Deciphering the origin of shells of NGC\,474 with the help of MOND}
\label{sec:mondpot}

MOND is a class of modified-gravity and inertia theories  \citep{milg83a,famaey12}. They all usually give similar predictions. Here, we assumed the QUMOND modified-gravity version of MOND \citep{qumond}. MOND has already proven to work well in virtually all types of galaxies \citep{gentile11,milg13,anddwarfii,samur14,mcgaugh16,lelli17}, with the possible exception of galaxies in clusters or slow rotators \citep{richtler08,rong18,bil19,tian20}. Nevertheless, NGC\,474 is a fast rotator residing in a poor group. The existing tests came out well for MOND in similar objects \citep{milg12,lelli17,rong18,bil19,shelest20}.

We derived the gravitational field of NGC\,474 by exploiting the MOND prescription for spherically symmetric systems \citep{qumond},
\begin{equation}
    a = a_\mathrm{N}\nu\left(a_\mathrm{N}/a_0\right). 
    \label{eq:mond}
\end{equation}
In this expression, $a_\mathrm{N}$ stands for the gravitational acceleration calculated in the usual Newtonian way from the distribution of baryons, $a_0$ is the acceleration constant of MOND, and $\nu$ is the interpolation function. In this paper, we assumed the so-called simple interpolating function,
\begin{equation}
    \nu(x) = \frac{1 + (1 + 4x^{-1})^{1/2}}{2},
    \label{eq:nu}
\end{equation}
and adopted $a_0 = 1.12\times 10^{-10}$\,m\,s$^{-2}$, consistent with recent measurements \citep{mcgaugh16}. 
Because of the proximity of the massive galaxy NGC\,470 to NGC\,474, the so-called external field effect of MOND should be considered. It is neglected in the main part of the paper, and in \app{other}, we show that including it does not change our conclusions substantially. In that appendix we also discuss how our results{ for MOND} change when we adopt an alternative, but less probable, stellar mass of the galaxy{, and we show that the estimates of the ages of the shell generations in NGC\,474 are not sensitive to the distribution of mass in the galaxy.}  

\subsection{Adopted shell identification and determination of shell ages}
\label{sec:ident}
{ The analytic model of evolution of shell radii in NGC\,474 for MOND is shown in} \fig{shcomp} by the curved lines. The radii of the observed shells are indicated by horizontal lines. 
We found the following shell identification to be most consistent with the data (we discuss another distribution that we found less consistent in \app{alternative}). The shell system was formed in at least two generations. The first and second generation consist of stars released from the secondary at look-back times $T_{\rm 1G} = 1280$\,Myr and $T_{\rm 2G} = 870$\,Myr during two pericentric passages of the galaxies. These times are marked in \fig{shcomp} with vertical dashed lines. The shells that{ were assigned to} the first generation are drawn in red, and the shells formed in the second generation are drawn in green. The assigned identification of each observed shell is indicated in \tab{shtab}. The Arabic numerals stand for the serial numbers, and the Roman numerals show the generation numbers. The advantage of this shell identification is that the model reproduces the observed shell radii very precisely. As we show in \sect{stelpop}, it also explains the ages of the young stellar populations in the galaxy well.

\subsection{MOND simulation of the formation of shells in NGC\,474}
\label{sec:sim}

\begin{figure*}
        \centering
        \includegraphics[width=17cm]{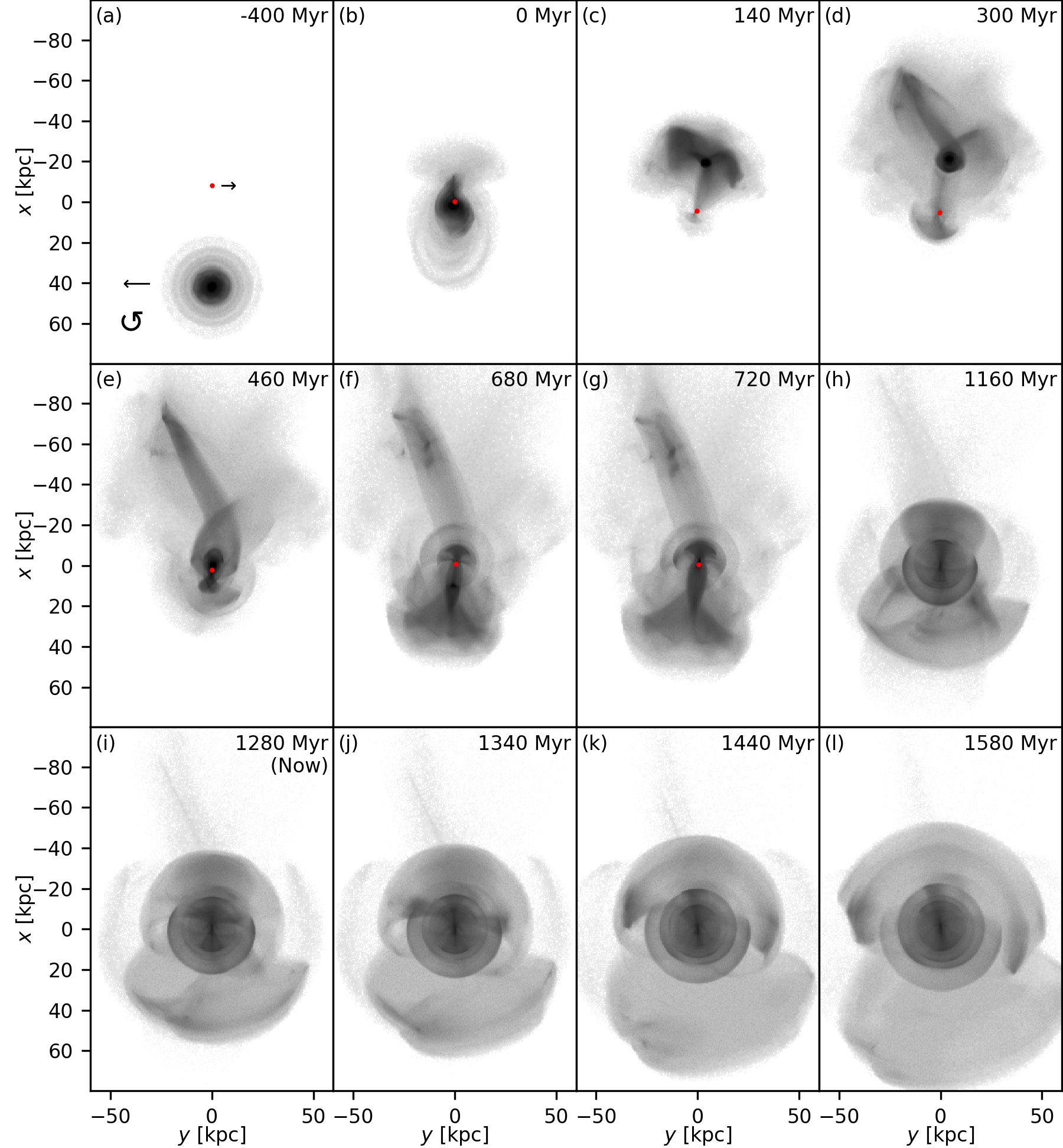}
        \caption{Snapshots of the stellar surface density from the MOND simulation. The numbers in the panels indicate the time after the first pericenter of the galaxies.  Only the material that originally belonged to the secondary galaxy is displayed. The position of the primary is marked by the red point in the first several plots. At later times, it virtually lies at the origin. The analytic models predict that the shell radii of NGC\,474 are reproduced at the time corresponding to panel (i).} 
        \label{fig:snap}
\end{figure*}

\begin{figure*}
        \centering
        \includegraphics[width=17cm]{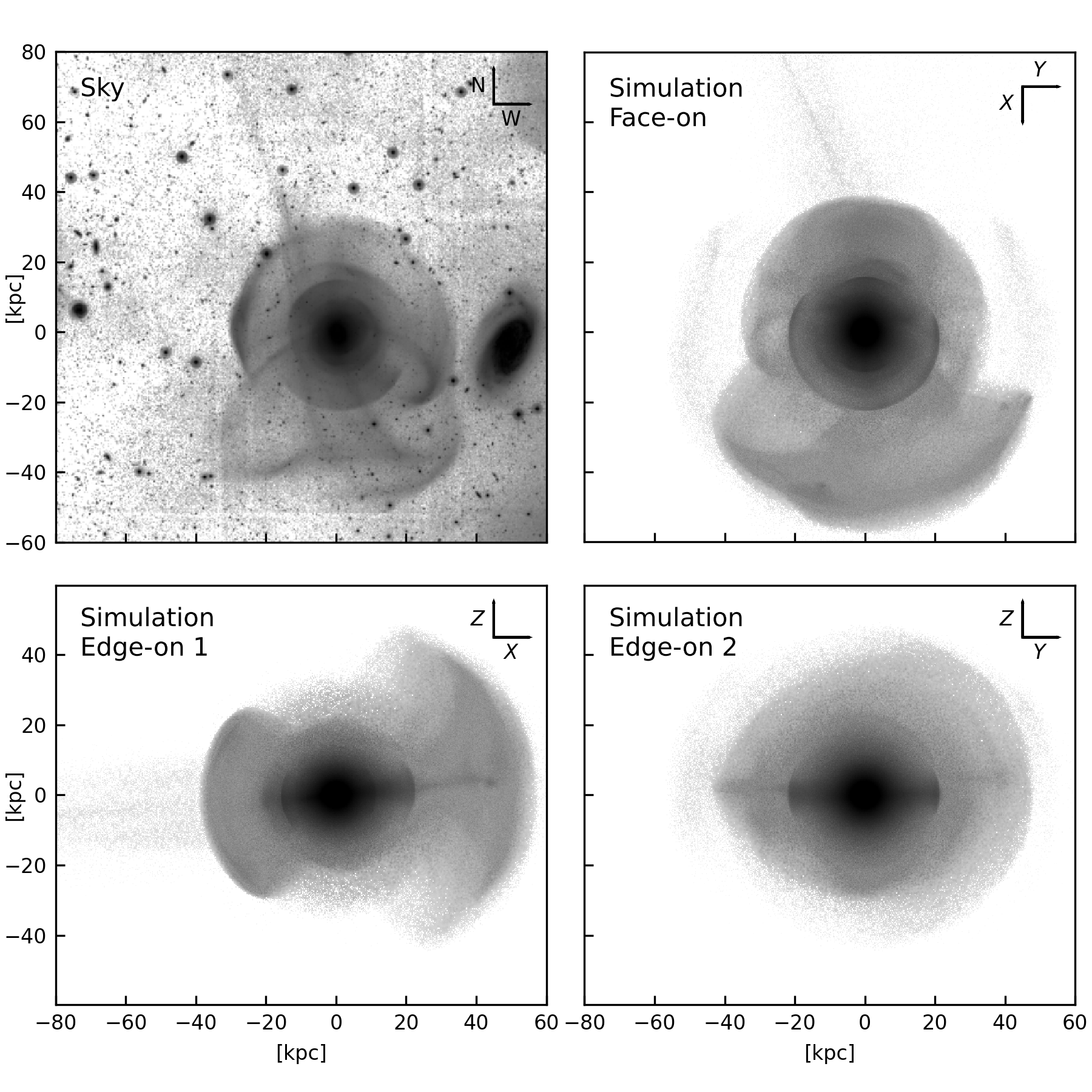}
        \caption{Morphology of the observed and simulated galaxy in the MOND simulation. Top left: Real galaxy (image taken in the MATLAS survey). 
        Top right: Simulated galaxy viewed perpendicular to the plane of collision and to the original plane of the disk of the secondary galaxy. 
        The morphology of the real galaxy is reproduced best by the simulation from this angle of view. Bottom left and right: Two perpendicular views of the simulated NGC\,474 in the plane of the sky.     
        The orientation of the simulation axes is indicated in each panel by the compass. We show the simulation at the time of 1280\,Myr after the first pericentric passage of the merged galaxies, i.e., at time \tig, when the analytic model of shell propagation predicts that the observed shell radii should be reproduced. The images of the simulated galaxy capture the stellar particles of the primary as well as the secondary galaxy.{ The secondary galaxy in the simulation originally approached the primary from the positive part of the $x$-axis.}
        }
        \label{fig:morph}
\end{figure*}

{ Here we describe our best MOND simulation.  We set the mass of the accreted spiral secondary galaxy to be one-sixth of the mass of the elliptical primary, in agreement with the observational constraints (\sect{obs}).  We thus explored various orbital configurations of the merger in order to reproduce the observed morphology of the tidal features  at time \tig \ after the first pericenter, as indicated by the results of the shell identification method  (\sect{ident}).   We strove primarily to reproduce the morphology of shells $A$ and $B$ and of the northern stream.  See \app{details} for the details of the way we used to find the initial conditions.   A nearly radial but mildly retrograde orbit of the secondary is necessary.} The orbital plane coincided with the plane of the sky and the spin vector of the secondary pointed directly to the observer. By matching the orientation of the tidal features in the simulated galaxy to the observations, we found that the secondary approached NGC\,474 from the south.

The progress of the simulated collision is depicted in \fig{snap}\footnote{A video showing the simulation can be found at \url{http://galaxy.asu.cas.cz/~bilek/col3MOND6-pubblend.mp4}.}. The times in the panels indicate the time after first pericenter. Only the stars that originally belonged to the secondary are shown. The position of the center of mass of the primary is marked by the red point in the first several panels. In the remaining panels, the center lies virtually at the origin of the coordinate system.  Panel (a) depicts the initial state of the simulation\footnote{The rings in the outskirts of the secondary are caused by insufficient virialization of the secondary in this region. Because the region contains only a small fraction of the mass of the secondary, we do not expect any substantial influence on the outcome of the simulation.}. We indicate the velocity vectors of the galaxies\footnote{The relative velocity of the galaxies was very low, such that the collision was nearly radial.} and the sense of rotation of the secondary. The galaxies started with a zero relative radial velocity and a low tangential velocity. Panel (b) captures the first pericenter passage. The pericentric distance was 2\,kpc and the relative velocity 390\,km\,s$^{-1}$.
In panels (c) and (d), the material brought in by the secondary mostly recedes from the primary, with the exception of the particles that became most tightly bound to the primary and therefore have the shortest oscillation periods.{ A few of them, but more of them in panel (d)}, are already in their second apocenter with respect to the primary, forming the first shell. In panel (e), the surviving core of the secondary moves through its second pericenter with respect to the primary. The time difference between the pericentric passages was 460\,Myr, which agrees well with what the adopted shell identification implies, that is, $\tig-\tiig= 410$\,Myr. In panels (f) and (g), tidal dwarf galaxies form in the tail in the upper part of the system. The largest shell in the bottom half of the galaxy shows a clumpy structure until about the time depicted in panel (h). The time for which the analytic model predicts the reproduction of shell radii in NGC\,474, \tig, corresponds to  panel (i). In panels (j) and (k), the largest shell in the upper half of the galaxy contains substructures that resemble{ the radial features} seen in shell $B$ of the observed galaxy. The simulation was terminated at the state shown in panel (l). 

The simulation snapshots also show us why the shells of NGC\,474 are so bright. At earlier times, before \tig, the shells are not developed well. The bright tidal features do not have the typical morphology of a shell. At later times, as the galaxy virializes toward equilibrium, the contrast of the tidal features decreases. We currently observe the shells at the optimal age.

We compared the observed galaxy with the simulation at a time of \tig=1280\,Myr after the first pericentric passage, when the analytic models predict that the observed shell radii should be reproduced (\sect{ident}). The comparison is presented in the top two panels of \fig{morph}. We note many similarities. The radii of several observed shells were replicated. This is a sign that the analytic models of shell evolution work well (see also \app{ansim}). It is reassuring that a radial stream formed at the place of the northern stream in NGC\,474. The northern stream of NGC\,474 is connected to the largest shell in the galaxy, $A$, which lies on the opposite side of the galaxy from the stream. An indication of this is seen for the simulated stream. The connection was more apparent in some of the other simulations we explored. The real shell $A$ is clumpy, while the simulated shell is rather smooth at time $T_{\rm 1G}$ after the merger. It contained distinct substructures only about 100\,Myr sooner before that time (panels (f)-(h) in \fig{snap}), however. The simulation suggests that the irregularity of the shell is a residue of the chaotic structure of the secondary after the first passage through the center of the primary. Shells with higher serial numbers appear smoother, likely because of the gradual stretching of the accreted material in the phase space \citep{quinn83,quinn84,bilcjp}.
The azimuthal extent of the two largest shells in the simulation is similar to that of shells $A$ and $B$ in NGC\,474. The second largest shell above the center of the simulated galaxy appears deformed, such that it deviates from a  circular shape. This is also the case of the observed shell $C,$ which lies at a very similar position. The observed shell $C$ lies almost at the same distance from the center of NGC\,474 as shell $D$. For the deformed simulated shell, there is also a shell at the opposite side of the galaxy that has nearly the same radius; see \app{ansim}. They are{ most probably} the same shell encircling the galaxy, with a brightness that varies in azimuth.  The loop in the left part of the largest simulated shell resembles one of the loops in the observed shell $A$.  The second largest simulated shell contains a hook-like structure that resembles the streams at the southern edges of the observed shell $B$.  \fig{snap} shows that the streams become more developed at later time steps of the simulation. We point out that the azimuthal extent and the internal morphology of the shells in the simulation are highly sensitive to the structure of the secondary used in the simulation (\app{details}). There is still much room for improvement of the agreement of the simulation with observations through an exploration of the parameter space.  For example, the faint wings of the simulated galaxy at $x \approx 0\,$kpc and $|y|\approx 50\,$kpc were formed only for some of the models of the secondary with which we experimented. Nevertheless, the simulations suggest that the {wings} could be in NGC\,474, and in that case, they should be only slightly dimmer than the northern stream in their brightest northern parts.{ An image that would reach only a slightly deeper surface brightness limit than the MATLAS image might capture them.}

In addition, the simulation agrees with the observed kinematics of NGC\,474.
\citet{fensch20} measured line-of-sight velocities of shell $B$ and of several globular clusters and planetary nebulae associated with it. Most of the measurements lie within a narrow range of $\pm20$\,km\,s$^{-1}$. The radial velocity of the center of the galaxy falls in the same range. \citet{alabi20} measured the stellar velocity dispersion of the shell as $18\pm9$\,km\,s$^{-1}$. In the simulation, we selected a spot in the galaxy that according to the morphology corresponds to the spot in the real galaxy probed by the observational works and inspected the kinematics of the particles in it. We found that the simulated shell $B$ has the same radial velocity as the center of the galaxy, and its velocity dispersion is 21\,km\,s$^{-1}$. This agrees well with the observations.

It would be interesting to know how NGC\,474 looks when it is viewed from different directions. The simulation can give us an approximate answer.  The bottom two panels of \fig{morph} show two views of the simulated galaxy that are perpendicular to the observed line of sight. They are edge-on views with respect to the secondary galaxy before the merger.  The figures demonstrate that the azimuthal extent of shells depends on the viewing angle. 
The bottom right panel also shows two shells at $|y|=20\,$kpc that lie at opposite sides of the galaxy, but have the same radii. Such pairs of shells are observed in some galaxies. Compared with the top right panel, the two shells at least in this simulation are actually a single shell with a large opening angle that is seen from a suitable point of view. The viewing angle can therefore influence the number of observed shells (see also \citealp{bil16,mancillas19}).

\begin{figure*}
        \centering
        \includegraphics[width=17cm]{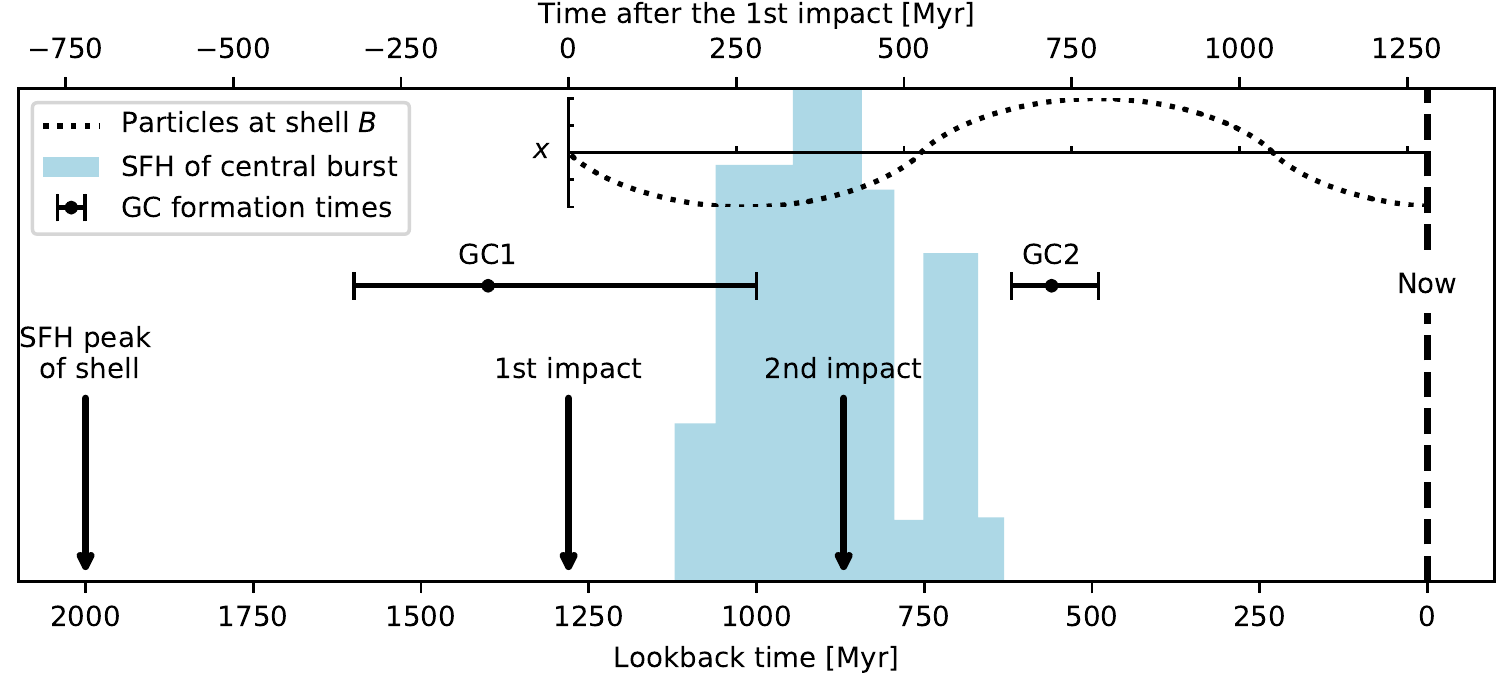}
        \caption{Timeline of the formation of NGC\,474 according our scenario. The arrows indicate the peak of the star formation history of the stellar population of shell $B$ \citep{alabi20} and the first and second pericentric passage of the secondary. The star formation history of the young stellar population \citep{fensch20} is shown by the light blue bar plot{ (in arbitrary units)}. The horizontal error bars mark the ages of the young globular clusters GC1 and GC2 \citep{fensch20}.{ The dotted line represents the  evolution of the $x$-coordinate of a particle that was released from the secondary during the first pericentric encounter of the galaxies and that currently lies at the edge of the shell $B$, assuming that the merger was radial. The $x$-coordinate here coincides with the collision axis, so that the secondary galaxy approached from the upper part of the $x$-axis.}  } 
        \label{fig:events}
\end{figure*}

\subsection{Nature of individual shells in NGC\,474}
\label{sec:nature}
{ Having at our disposal the shell identification and the simulation, we comment here on the nature of a few special shells in NGC\,474.} The irregular structure $A$ is assigned shell 1 of the first generation. This interpretation is also supported by the simulation we presented in \sect{sim}. In the simulation, shell 1 of the first generation is joined with a stream on the opposite side of the galaxy. A similar configuration is also seen in the real galaxy: shell $A$ is joined with the northern stream of NGC\,474. In the simulation, the counterpart of shell $A$ was clumpy{ somewhat before the time $T_{\rm 1G}$.}

The model of the shell propagation suggests that shell $B$ is the second shell of the first generation.
It might also be a blend of the second shell of the first generation and the first shell of the second generation. The image of the shell does not show any convincing indication that it is a blend.
On the other hand, in the simulation, shell $B$ has a clear counterpart that is a blend.

Shells $E$ and $F$ are considered a single shell{ in our identification}. Based on the image of the galaxy, it is not obvious whether structures $E$ and $F$ are connected. They could be two parts of one shell whose surface brightness varies in azimuth. This phenomenon is seen in shell $B,$ and we found signs of this phenomenon in the simulation as well (\app{ansim}). In addition, there is a stellar stream in the galaxy that is overlaid on the region in which we would expect the connection between shells $E$ and $F$.{ It is plausible that the stream might hide the connection.} Similarly, it is possible to identify shells $C$ and $D$ as one shell encircling the galaxy.

\subsection{Explaining the ages of {the young} stellar populations in NGC\,474}
\label{sec:stelpop}
In \fig{events} we compare the ages of the shells of NGC\,474 derived in \sect{ident} with the ages of various stellar populations observed spectroscopically by \citet{fensch20} and \citet{alabi20}. We first discuss the mass-weighted star formation history of the young stellar population in the center of NGC\,474 derived by \citet{fensch20}. It is shown in{ arbitrary units} in \fig{events} by the bar plot.  The figure indicates that the stars started to form after the first encounter of the primary and secondary. This timing appears logical. Hydrodynamic simulations of the formation of shell galaxies show that the gas brought in by the secondary settles quickly in the center of the primary, leading to a central starburst \citep{hernquist92, weil93}. This is because the material in any small part of a shell moves at the same time toward and away from the center of the host galaxy. This is not possible for gas clouds. They collide with each other, dissipate energy, separate from the accreted stars, and settle in the center of the galaxy. Some delay between the first pericentric passage of the primary and secondary and the onset of the central starburst is expected. The reason is that shortly after the first pericenter of the primary and secondary, all the material brought by the secondary still moves in the same direction as before the collision. The core of the secondary is accompanied by two tidal tails, one preceding the core and one following it (\fig{snap} tiles (c) and (d)). The trailing arm is the first to reverse direction and fall back onto the center of the primary. When this material changes the direction of its motion again, the first shell is formed, and the paths of the particles of the accreted material start to cross each other (\fig{snap} (d)). This is when the central starburst can begin.  After the trailing arm has fallen back onto the center of the primary, it is followed by the core of the secondary (\fig{snap}(e)) and then by the leading arm (\fig{snap} panels (f) to about (h)). This explains why in  \fig{events}  the second impact of the secondary occurs approximately in the middle of the star-forming period.  The star formation history shows an indication of a secondary peak at a look-back time of about 700\,Myr, which might have been caused by the second pericentric passage of the galaxies. If any star formation occurred in the secondary galaxy before the first impact, these stars, as noncollisional objects, would spread throughout the whole extent of the shell system and would not settle preferentially in the center of the primary. Moreover, for a  mildly retrograde encounter, which our simulations suggests has created the shells, we do not expect a substantial increase in star formation activity before the first encounter because retrograde encounters are known to produce much weaker deformations of the galaxies than prograde encounters in the initial stages of the merger \citep{mo10}.

\citet{fensch20} also measured the ages of two young globular clusters lying close to the edge of shell $B$. Their positions are marked in the image of the galaxy in \fig{annot}. The age of GC1 is consistent with the first encounter of the galaxies, as evident from \fig{events}. This suggests that the cluster was formed during the tidal shock within the secondary during the pericentric passage. The star formation timescales of globular clusters can be as short as a few million years \citep{puzia06}. 

The age of GC2 does not agree well with the time of the first or second pericentric passage of the galaxies. To elucidate its formation, we integrated the positions of particles that currently lie at the edge of shell $B$  backward
in time, assuming that it is the second shell of the first generation\footnote{The serial number sets the current radial velocity of the particles at the edge of the shell.} and that the particles move on a radial trajectory along a coordinate axis. The resulting trajectory is shown in \fig{events}{ in the inset plot as the dashed black sinusoid. It corresponds to the trajectory of the cluster or its progenitor gas cloud}.  The comparison of the trajectory to the other events in the galaxy yields the following formation scenario of GC2. First, the secondary passed through its first pericenter with respect to the primary. This caused the progenitor gas cloud of GC2 to detach from the secondary. Subsequently, the secondary impacted the center of the primary for the second time, being followed by the progenitor cloud. The cloud formed GC2{ about 250\,Myr} after this, perhaps because of the tidal shock because of a collision with some of the other gas clouds that were concentrated in the central region.{ It is also notable that the time of the formation of the cluster is consistent with the moment of apocenter of its progenitor gas cloud. This indicates that the two events might be related. For example, the parts of the cloud that passed through the apocenter of their orbits might collide with the parts of the cloud that still receded from the center of the galaxy, which caused the cloud to collapse. We probably need hydrodynamical simulations in order to explain how the young globular clusters formed. }

\citet{alabi20} reconstructed the star formation history of the stellar population of shell $B$ and found that the star formation increased gradually until about 2\,Gyr ago and then dropped quickly. The authors suggested that the shell-forming merger induced the peak of star formation, and then the disruption of the secondary led to the quenching of the star formation. This dating of the merger is not consistent with our scenario. On the other hand, the MaNGA survey reported the same shape of star formation histories for all galaxies in the mass range $10^9-10^{10}\,M_\sun$ \citep{sanchez19}, which coincides with  the supposed pre-merger mass of the secondary. We therefore consider the observed shape of star formation history of the shell to be unrelated to the shell-forming merger.

\subsection{MOND indicates that NGC\,470 initiated the formation of the shells in NGC\,474}
\label{sec:init}

In this section, we point out that there are indications that the merger that formed the shells in NGC\,474 was started by a three-body encounter of NGC\,474, the accreted galaxy, and the neighboring galaxy NGC\,470.{ This option appears quite convincing in the context of MOND.} In restricted three-body simulations of{ shell-forming mergers} \citep[e.g., ][]{ebrova12}, it is observed that when the primary and secondary galaxies are released with zero relative velocity, the apocenters near the galactocentric radius of most particles accreted onto the primary are equal to the inital distance of the secondary. This is because of the conservation of energy. In these simplified simulations, dynamical friction and self-gravity of the secondary are neglected. Nevertheless, in MOND,  dynamical friction during galaxy mergers is known to be low compared to{ simulations with Newtonian gravity and dark matter} \citep{tiret08,renaud16}. From this point of view, it is striking that NGC\,474 hosts very prominent shells up to a galactocentric distance of about 30\,kpc, but lacks any substantial structures at higher distances. It seems that the secondary impacted NGC\,474 with a relatively low velocity, as if it fell from rest from a distance of about 30\,kpc. This was confirmed by our self-consistent MOND simulations, in which we obtained the best match of the tidal structures when the galaxies fell onto each other nearly from rest from a distance of 50\,kpc. This raises the question of how the secondary could have appeared a few tens of kiloparsec away from NGC\,474 while being nearly in rest with respect to it. A possible solution is that the two galaxies appeared in this configuration because of a three-body interaction with NGC\,470. It is desirable to verify this option by simulations.

Observational clues also support this scenario. Galaxies NGC\,474 and NGC\,470 are embedded in the same HI cloud, and NGC\,470 shows structures resembling tidal arms pointing to NGC\,474 \citep{schiminovich97, rampazzo06}, as if they had a past encounter and exchanged some material \citep{cullen06,bil18,young20}. In addition, the deep MATLAS image of NGC\,470 reveals that the outer parts of the galaxy are asymmetric, which might be a result of past tidal interaction\footnote{In MOND, tidal features can also form by nonmerging galaxy flybys \citep{bil18,bil19b}. In the case of NGC\,474, the flyby of NGC\,470 comes to mind. Nevertheless, we consider here the minor merger scenario because it seems to explain all constraints well and because it is questionable whether the mass transfer mechanism can produce shells with the morphology observed in NGC\,474.}.

\begin{table}
\caption{Estimates of the ages of shell generations in NGC\,474 for various dark matter halos.}            
\label{tab:dmmod} 
\centering                                  
\begin{tabular}{ccllll}          
\hline\hline                       
\multicolumn{2}{l}{Name or} &
 $\log_{10}\frac{M_\mathrm{vir}}{M_\sun}$ & $r_\mathrm{s}$ & \tig & \tiig\\ 
SHMR $\frac{\Delta}{\sigma}$ & HMCR $\frac{\Delta}{\sigma}$  & & [kpc] & [Myr] & [Myr] \\
\hline   
-1 & -1  & 11.88 & 32.9 & 1410 & 990 \\
-1 & 1 & 11.88 & 15.8 & 1160 & 800 \\
1 & -1 & 12.54 & 65.1 & \makecell[l]{1080\\(1560)} & \makecell[l]{860\\(1090)}\\
 1 & 1 & 12.54 & 31.2 & \makecell[l]{810\\(1150)} & \makecell[l]{650\\(810)}\\
\multicolumn{2}{l}{Preferred halo}  & 12.21 & 32.0 & 1050 & 810\\
\multicolumn{2}{l}{MOND-fitting halo} & 11.8 & 17.9 & 1280 & 870\\
\hline                                             
\end{tabular}
\end{table}

\begin{figure*}
        \centering
        \includegraphics[width=17cm]{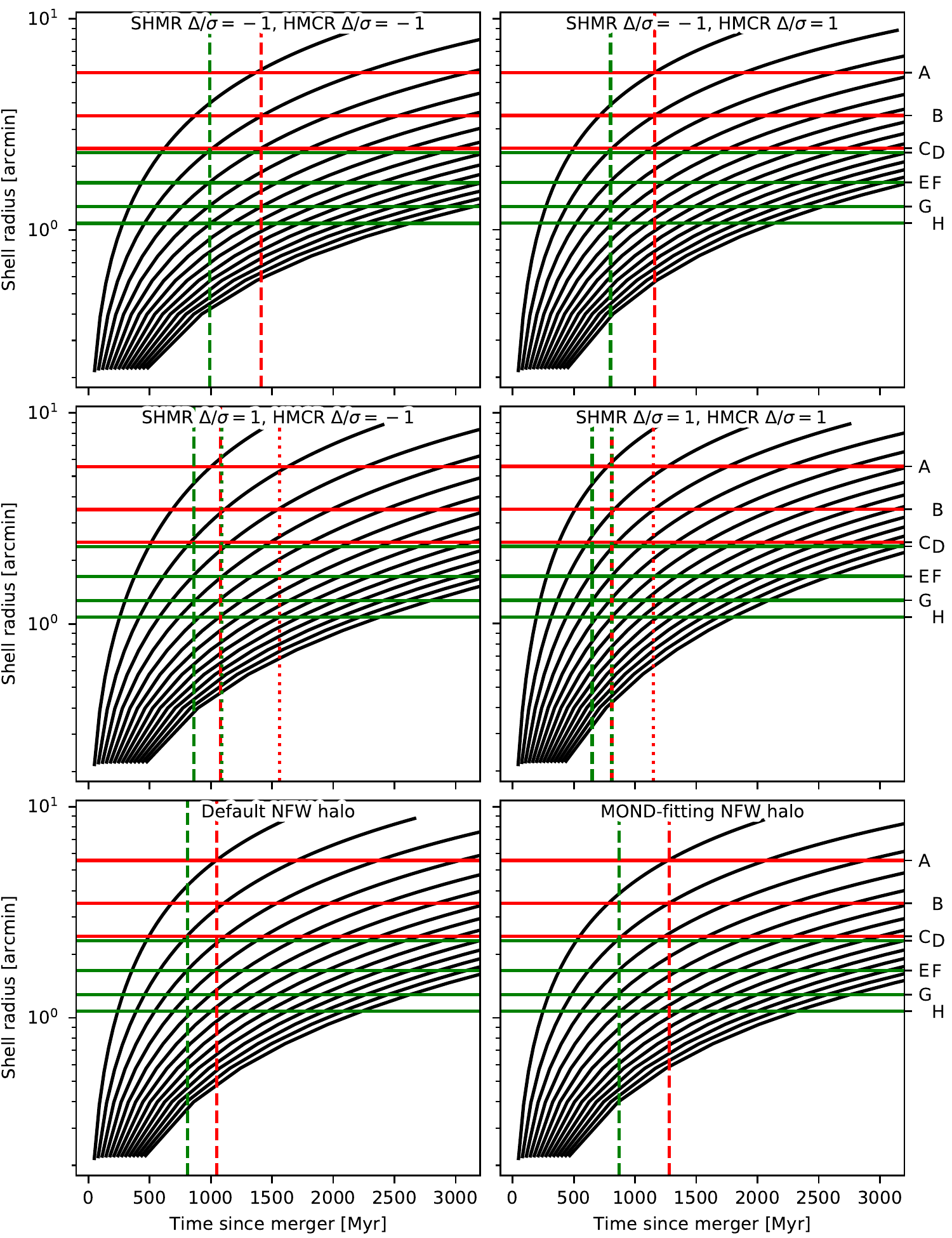}
        \caption{{ Comparison of  the observed shell radii in NGC\,474 with  the modeled evolution of shell radii for Newtonian gravity and various choices of dark matter halos. Top two rows: Illustration of the influence of the uncertainty arising due to the intrinsic scatter of the SHMR and HMCR. Bottom left: Models of shell evolution for the preferred dark matter halo. Bottom right: Models of shell evolution for the dark matter halo that lends the galaxy a similar gravitational potential as MOND. }}
        \label{fig:shcompNFW}
\end{figure*}

\begin{figure*}
        \resizebox{\hsize}{!}{\includegraphics{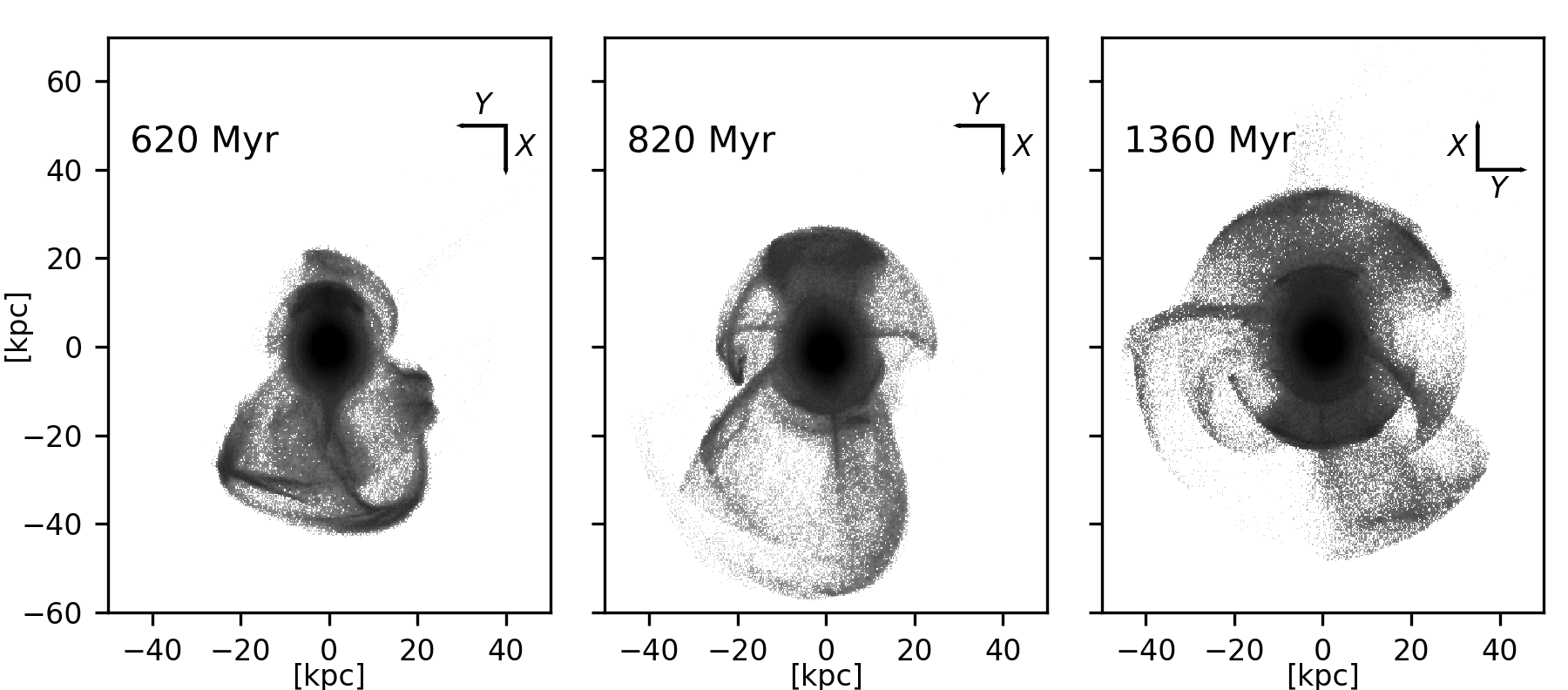}}
        \caption{Snapshots from the Newtonian simulation. They were chosen such that they reproduce the morphology of some of the features in NGC\,474 well.
        }
        \label{fig:dmsim}
\end{figure*}

\section{Analysis with Newtonian gravity}
\label{sec:nwt}
\subsection{Shell identification for various dark halos}
\label{sec:dmpot}
{When Newtonian gravity is assumed, galaxies must be surrounded by dark matter halos. The distribution of dark matter cannot be predicted uniquely from the distribution of the observable matter. Nevertheless, we can estimate the parameters of an assumed profile of the dark matter halo from their scaling relations with the stellar mass of the galaxy. We  considered the stellar to halo-mass relation (SHMR) of \citet{behroozi13} and the halo mass\,--\,concentration relation (HMCR) of \citet{diemer15}. We also took from these works the intrinsic scatters of the relations. We  assumed that the halos have Navarro-Frenk-White (NFW)  profiles \citep{nfw}. The analytic models below are calculated for gravitational potentials generated by NFW halos and the observed distribution of stars of NGC\,474 (\sect{obs}). The masses and scale radii of the dark halos considered in this section are summarized in \tab{dmmod}, along with the ages of shells that the shell identification method prefers for them.

The top two rows of panels in \fig{shcompNFW} demonstrate how the intrinsic scatter in the dark halo scaling relations influence the evolution of shell radii according to the analytic models. The parameters of the halos were set to deviate from the mean scaling relations either by plus or minus one standard deviation of the intrinsic scatter of the relations, as indicated in the figure. All these dark matter halos are well possible for NGC\,474. The two-generation shell identification stated in \tab{shtab}, which was inspired by the MOND model,  works also for these four Newtonian models. The times corresponding to the creation of the two generations of shells are marked in \fig{shcompNFW} by the vertical dashed red and green lines.  For the two models shown in the second row of \fig{shcompNFW}, another shell identification is possible. The times corresponding to the first and second generation are marked by the red and green vertical dotted lines, respectively, and are stated in parentheses in \tab{dmmod}. The grouping of the observed shells in the first and second generation is the same as before. In total, the shell radii in NGC\,474 are consistent with the standard cosmology, as far as we can tell from the analytic models. Nevertheless, the choice of the dark matter halo has a strong influence on the ages of the shell generations and thus on the times since the pericentric passages of the primary and the secondary. With the considered choices of halo parameters, the first encounter might have occurred between about 810 and 1410\,Myr ago, if the shell identification from \tab{shtab} is correct. With the other shell identification, the time since the formation of the first generation might even be 1560\,Myr. These uncertainties prevented us from making a detailed comparison of the analytic models with the ages of the young stellar populations, like we did for MOND.

If we rely on the validity of HMCR and SHMR, then the halo that follows the mean HMCR and SHMR is most probable and therefore preferred. The preferred halo has a scale radius of 32.0\,kpc and a virial mass of $10^{12.21}\,M_\sun$. The resulting analytic model of the shell evolution in NGC\,474 is shown by the curved lines in the bottom left panel of \fig{shcompNFW}. Either the whole shell system was formed in a single generation 1050\,Myr ago, or there are two shell generations in the galaxy, the first 1050\,Myr old and the second 810\,Myr old.
 
As noted in \citet{bil19}, for giant early-type galaxies, it is possible to choose an NFW halo such that the Newtonian gravitational field of the galaxy mimics the MOND gravitational field at high precision. For NGC\,474, we found an NFW halo such that the Newtonian gravitational acceleration differs from the MOND gravitational acceleration by less than 5\% within a sphere of the radius of the outermost observed shell, that is, 50\,kpc. The halo is characterized by a  scale radius of 17.9\,kpc and a virial mass of $10^{11.81}\,M_\sun$. Its mass deviates from the mean SHMR downward by 1.2 standard deviations of the intrinsic scatter of the relation, and the concentration of the halo follows the mean HMCR nearly exactly. Even this halo is thus well consistent with  the standard cosmology. The modeled evolution of shell radii in this gravitational potential, shown in the bottom right panel of \fig{shcompNFW},  is virtually the same as for the MOND potential. This demonstrates that interpretation of the young stellar populations that we performed for MOND is also possible for Newtonian gravity, at least as long as we consider only the analytic models of shell evolution. Differences between MOND and Newtonian gravity might result from other effects that the analytic models currently cannot take into account, such as the expected difference in dynamical friction in the two types of gravity \citep{bildiz,vakili17}. Finally, we point out that the results of this section imply that  the shell identification from \tab{shtab} is not specific to MOND alone; it is also acceptable for a wide range of realistic Newtonian potentials of NGC\,474.}

\subsection{Newtonian simulation}
\label{sec:nwtsim}
{ We also performed simulations with Newtonian gravity. The purpose was to demonstrate that even with the standard gravity, it is possible to form shells similar to those in NGC\,474.}

The details of the most successful Newtonian simulation\footnote{A video showing the simulation can be found at \url{http://galaxy.asu.cas.cz/~bilek/col3IC2-2-pubblend.mp4}.} that we describe below can be found in \app{details}.{ The galaxies were assigned the preferred dark matter NFW halos, as explained in \sect{dmpot}. The primary and secondary galaxies were released with zero relative radial velocity from a relative distance of 100\,kpc, approaching on a  nearly radial orbit}. A few interesting moments in the simulation are depicted in  Fig.~\ref{fig:dmsim}. The indicated time is counted since the first pericentric passage of the galaxies. In the left panel, the largest structure at the bottom of the galaxy resembles shell $A$ in NGC\,474 in its internal structure and size. The simulated structure is shell 1 of the first generation. In the middle panel of \fig{dmsim}, the largest shell above the center of the galaxy resembles shell $B$ of NGC\,474 in its size and by being terminated at the bottom by the two hook-shaped radial streams. The largest shell below the galaxy again resembles the observed shell $A$ and is again the first shell of the first generation. In all Newtonian simulations we explored, there never formed a long stream similar to the northern stream in NGC\,474. A certain indication of it can be noted in the right panel of \fig{dmsim}, above the galaxy. This is the remnant of the structure that we considered as the counterpart of shell $A$ in the previous panels. The shape and substructure of the largest shell above the center of the galaxy resembles shell $B$ of NGC\,474. The sizes of the three largest shells also match the sizes of the observed shells well.  

We point out that we explored only a small part of the parameter space for both MOND and Newtonian simulations. It would very likely be possible to improve the match of the morphologies of the simulated and observed tidal structures by tuning the pre-merger profiles of the galaxies and of the orbital configuration. For this reason, it is not possible to distinguish between Newtonian and MOND gravity on the basis of the simulations presented in this paper. The expected differences in the formation of shells with Newtonian and MOND gravity were discussed in \citet{bildiz} and \citet{vakili17} on the basis of analytic arguments.

\subsection{No need for a three-body encounter with Newtonian gravity}
\label{sec:nwtinit}
{ In the context of Newtonian gravity, we can still argue for an initiation of the merger by a three-body encounter like in \sect{init} because of the distribution of gas around NGC\,474 and NGC\,470 and the optical irregularity of the outer parts of NGC\,470. However, these are not necessarily related to the shells in NGC\,474. In the Newtonian case, we lack our main argument for the three-body interaction from \sect{init}. In MOND, it should be difficult to produce the relatively small shells like in NGC\,474 without a three-body encounter. Our Newtonian simulations rather showed the opposite problem: it seems difficult to produce a system of tidal features that is at least as large as observed (no counterpart of the northern stream), even if the secondary falls from a very large distance. In the context of Newtonian gravity, we thus found only ambiguous evidence for the scenario of the three-body interaction.}

\section{Summary and conclusions}
\label{sec:sum}
{ The galaxy NGC\,474 hosts numerous exceptionally bright stellar shells and other tidal features. Prompted by the recent spectroscopic observations of this prominent galaxy \citep{alabi20,fensch20}, we attempted to give additional constraints on its formation. We applied the analytic shell identification method \citep{bil13} on the measured radii of the shells in the galaxy. This helped us to elucidate the nature of many of the tidal features in the galaxy and to deduce the time since the shell-forming merger.  
In addition, we made self-consistent $N$-body simulations of merging galaxies that produced galaxies with tidal features similar to those in NGC\,474 (Figures~\ref{fig:snap}, \ref{fig:morph}, and~\ref{fig:dmsim}). The simulations and particularly the analytic models helped us to explain the link between the merger and the young stellar population discovered in the galaxy by the spectroscopic studies. We made the analysis assuming either the standard Newtonian gravity or the MOND gravity.} 

{ Assuming MOND, we arrived at }the following scenario of the formation of the tidal structures in NGC\,474 that seems to explain the observed properties of NGC\,474 and its neighbor NGC\,470. NGC\,474 was hit about 1.3\,Gyr ago by a small spiral galaxy with a stellar mass of about $10^{9.8}\,M_\sun$. The accreted galaxy approached NGC\,474 in the plane of the sky from the south, with its rotation axis aligned with the line of sight. The merger orbit was nearly radial but mildly retrograde, such that the pericenter was located east of the center of NGC\,474. The merger had a relatively low orbital energy. There are indications that it was initiated by a three-body encounter of NGC\,474, the accreted companion, and the nearby massive spiral NGC\,470. The tidal shock experienced by the secondary during the first pericentric approach led to the formation of the young globular cluster GC1 observed by \citet{fensch20}. The secondary galaxy was partly disrupted, but its core stayed at least partly self-bound. 
Shortly after the first pericentric approach, the core developed massive leading and trailing tidal arms. The whole structure initially moved away from the core of the primary, but then it gradually started to fall back, initiating the oscillatory motion of the accreted material. The trailing arm was the first to fall back. As the paths of the accreted gas clouds started to cross each other, the gas separated from the accreted stars and settled in the center of NGC\,474. This induced the period of central star formation detected spectroscopically by \citet{fensch20}. The surviving core of the secondary impacted the center of NGC\,474 approximately in the middle of this star formation period, about 0.9\,Gyr ago, and gave rise to a second shell generation in the galaxy. Then the leading tidal arm reached the center of the primary. It carried a gas cloud that collapsed into the cluster GC2, observed by \citet{fensch20},  0.56\,Gyr ago after its passage through the center of the primary galaxy. 

The times since the two pericenters were recovered using the shell identification method \citep{bil13}. The method also tells us the serial and generation numbers of the observed shells. They are  listed in \tab{shtab}. We identified the irregular structure south of the galaxy to be shell 1 of the first generation.  It seems that its irregularity was inherited from the chaotic appearance of the secondary galaxy just after its first pericentric passage through the primary galaxy. The simulation repoduced not only many morphological features in the galaxy, but also the velocity dispersion of one of the shells measured by \citet{alabi20} and \citet{fensch20}. In relation to the validity of MOND, it is encouraging that we were able to find a scenario that fits all available constraints well.

{ The precision of the analytic models of time evolution of shell radii on which the shell identification relies is not explored well. It would be desirable to test the precision against multiple $N$-body simulations. Here we used our MOND simulation to show that the analytic models of shell evolution work well for shells with a morphology as observed in NGC\,474. The deviation of the simulated and modeled shell radii was better than 0.1\,dex.}

{ With Newtonian gravity, the dark matter halo is uncertain. The recovery of the merger times is not unique, and therefore it is not possible to unambiguously link the young stellar populations to the merger. Nevertheless, from the empirical point of view, MOND has proved to work well in predicting gravitational fields of galaxies similar to NGC\,474 \citep{milg12,lelli17,rong18,bil19,shelest20}, and therefore the above MOND scenario is probable regardless of gravity model, as far a we can tell based on the analytic models of shell evolution. We indeed showed that the gravitational field of NGC\,474 predicted by MOND can be reproduced by assuming a suitable NFW halo whose parameters are consistent with the theoretical correlations.  The simulations with Newtonian gravity were also able to reproduce much of the observed morphology of NGC\,474. The initiation of the formation of shells in NGC\,474 by NGC\,470 is not excluded, but neither is it required, {under the assumption of Newtonian gravity}. }

{ Coming to the conclusions that follow from both Newtonian and MOND results,  all gravitational potentials explored in this work, which should cover most of the realistic potentials of NGC\,474, imply a time since the shell-forming merger shorter than 1.6\,Gyr. We propose that the drop in star formation rate 2\,Gyr ago that was detected in the reconstructed star formation history of the stellar population of one of the shells by \citet{alabi20} is a consequence of secular evolution of the progenitor of the shells prior the merger, as supported by the observed star-forming histories in galaxies similar to the probable progenitor \citep{sanchez19}.}

{ The MOND simulation convincingly reproduced the overall structure of the shell system with the northern stream for the sizes and locations of the features. The Newtonian simulation was more successful in reproducing the internal structure of the two largest shells. This suggests that all tidal features observed in NGC\,474 originate from one accreted disk galaxy. There is still much room for improvement of the simulations because we did not explore the whole parameter space. For example, the morphology of the shells  (angular extent, internal structure) turned out to depend sensitively on the initial structure of the accreted galaxy. For this reason, the current simulations do not allow discriminating between Newtonian and MOND gravities. The internal structure of the shells or the northern stream seem to be promising features on which to focus such studies.} 

{ We found that one of the reasons why the shells in NGC\,474 are so prominent is that {they are observed} at the optimal moment. Earlier, the tidal debris did not take up the typical morphology of shells.  Later, the tidal  will lose contrast and disappear as a consequence of dynamical virialization.}

As a general note, shell galaxies seem to be merger remnants whose formation history is particularly simple to recover. The analysis can be simplified by assuming the MOND gravity (unless we deal with galaxies in clusters or slow rotators, where the applicability of MOND remains unclear; \citealp{richtler08,samur14, rong18,bil19}). Shell identification method can be used to estimate the time since the encounter just from shell radii, but the result can be ambiguous. The degeneracy can be broken by deriving the star formation histories of the centers of shell galaxies, which can now be performed routinely.

\begin{acknowledgements}
IE and MB acknowledge the support from the Polish National Science Centre under the grant 2017/26/D/ST9/00449.

MB is thankful for the financial support by {\it Cercle Gutenberg}.

BF acknowledges financial support from the European Research Council (ERC) under the European Unions Horizon 2020 research and innovation programme (grant agreement No. 834148).

 
\end{acknowledgements}


\bibliographystyle{aa}
\bibliography{citace}

\begin{thebibliography}{115}
\expandafter\ifx\csname natexlab\endcsname\relax\def\natexlab#1{#1}\fi

\bibitem[{{Alabi} {et~al.}(2020){Alabi}, {Ferr{\'e}-Mateu}, {Forbes},
  {Romanowsky}, \& {Brodie}}]{alabi20}
{Alabi}, A.~B., {Ferr{\'e}-Mateu}, A., {Forbes}, D.~A., {Romanowsky}, A.~J., \&
  {Brodie}, J.~P. 2020, \mnras, 497, 626

\bibitem[{{Amorisco}(2015)}]{amorisco15}
{Amorisco}, N.~C. 2015, \mnras, 450, 575

\bibitem[{{Arp}(1966)}]{arp66}
{Arp}, H. 1966, {Atlas of peculiar galaxies}

\bibitem[{{Atkinson} {et~al.}(2013){Atkinson}, {Abraham}, \&
  {Ferguson}}]{atkinson13}
{Atkinson}, A.~M., {Abraham}, R.~G., \& {Ferguson}, A. M.~N. 2013, \apj, 765,
  28

\bibitem[{{Behroozi} {et~al.}(2013){Behroozi}, {Wechsler}, \&
  {Conroy}}]{behroozi13}
{Behroozi}, P.~S., {Wechsler}, R.~H., \& {Conroy}, C. 2013, \apj, 770, 57

\bibitem[{{B{\'\i}lek}(2016)}]{bildiz}
{B{\'\i}lek}, M. 2016, arXiv e-prints, arXiv:1601.01240

\bibitem[{{B{\'\i}lek} {et~al.}(2014){B{\'\i}lek}, {Barto{\v{s}}kov{\'a}},
  {Ebrov{\'a}}, \& {Jungwiert}}]{bil14}
{B{\'\i}lek}, M., {Barto{\v{s}}kov{\'a}}, K., {Ebrov{\'a}}, I., \& {Jungwiert},
  B. 2014, \aap, 566, A151

\bibitem[{{B{\'\i}lek} {et~al.}(2016){B{\'\i}lek}, {Cuillandre}, {Gwyn},
  {Ebrov{\'a}}, {Barto{\v{s}}kov{\'a}}, {Jungwiert}, \&
  {J{\'\i}lkov{\'a}}}]{bil16}
{B{\'\i}lek}, M., {Cuillandre}, J.~C., {Gwyn}, S., {et~al.} 2016, \aap, 588,
  A77

\bibitem[{{B{\'\i}lek} {et~al.}(2020){B{\'\i}lek}, {Duc}, {Cuilland re},
  {Gwyn}, {Cappellari}, {Bekaert}, {Bonfini}, {Bitsakis}, {Paudel},
  {Krajnovi{\'c}}, {Durrell}, \& {Marleau}}]{bil20}
{B{\'\i}lek}, M., {Duc}, P.-A., {Cuilland re}, J.-C., {et~al.} 2020, \mnras,
  498, 2138

\bibitem[{{B{\'\i}lek} {et~al.}(2015{\natexlab{a}}){B{\'\i}lek}, {Ebrov{\'a}},
  {Jungwiert}, {J{\'\i}lkov{\'a}}, \& {Barto{\v{s}}kov{\'a}}}]{bilcjp}
{B{\'\i}lek}, M., {Ebrov{\'a}}, I., {Jungwiert}, B., {J{\'\i}lkov{\'a}}, L., \&
  {Barto{\v{s}}kov{\'a}}, K. 2015{\natexlab{a}}, Canadian Journal of Physics,
  93, 203

\bibitem[{{B{\'\i}lek} {et~al.}(2015{\natexlab{b}}){B{\'\i}lek}, {Jungwiert},
  {Ebrov{\'a}}, \& {Barto{\v{s}}kov{\'a}}}]{bil15}
{B{\'\i}lek}, M., {Jungwiert}, B., {Ebrov{\'a}}, I., \& {Barto{\v{s}}kov{\'a}},
  K. 2015{\natexlab{b}}, \aap, 575, A29

\bibitem[{{B{\'\i}lek} {et~al.}(2013){B{\'\i}lek}, {Jungwiert},
  {J{\'\i}lkov{\'a}}, {Ebrov{\'a}}, {Barto{\v{s}}kov{\'a}}, \&
  {K{\v{r}}{\'\i}{\v{z}}ek}}]{bil13}
{B{\'\i}lek}, M., {Jungwiert}, B., {J{\'\i}lkov{\'a}}, L., {et~al.} 2013, \aap,
  559, A110

\bibitem[{{B{\'\i}lek} {et~al.}(2019{\natexlab{a}}){B{\'\i}lek},
  {Samurovi{\'c}}, \& {Renaud}}]{bil19}
{B{\'\i}lek}, M., {Samurovi{\'c}}, S., \& {Renaud}, F. 2019{\natexlab{a}},
  \aap, 625, A32

\bibitem[{{B{\'\i}lek} {et~al.}(2018){B{\'\i}lek}, {Thies}, {Kroupa}, \&
  {Famaey}}]{bil18}
{B{\'\i}lek}, M., {Thies}, I., {Kroupa}, P., \& {Famaey}, B. 2018, \aap, 614,
  A59

\bibitem[{{B{\'\i}lek} {et~al.}(2019{\natexlab{b}}){B{\'\i}lek}, {Thies},
  {Kroupa}, \& {Famaey}}]{bil19b}
{B{\'\i}lek}, M., {Thies}, I., {Kroupa}, P., \& {Famaey}, B.
  2019{\natexlab{b}}, arXiv e-prints, arXiv:1908.07537

\bibitem[{{Binney} \& {Tremaine}(2008)}]{bt08}
{Binney}, J. \& {Tremaine}, S. 2008, {Galactic Dynamics: Second Edition}

\bibitem[{{Bizyaev} \& {Mitronova}(2002)}]{bizyaev02}
{Bizyaev}, D. \& {Mitronova}, S. 2002, \aap, 389, 795

\bibitem[{{Burkert}(1995)}]{burkert95}
{Burkert}, A. 1995, \apjl, 447, L25

\bibitem[{{Caldwell} {et~al.}(2017){Caldwell}, {Walker}, {Mateo}, {Olszewski},
  {Koposov}, {Belokurov}, {Torrealba}, {Geringer-Sameth}, \&
  {Johnson}}]{caldwell17}
{Caldwell}, N., {Walker}, M.~G., {Mateo}, M., {et~al.} 2017, \apj, 839, 20

\bibitem[{{Canalizo} {et~al.}(2007){Canalizo}, {Bennert}, {Jungwiert},
  {Stockton}, {Schweizer}, {Lacy}, \& {Peng}}]{canalizo07}
{Canalizo}, G., {Bennert}, N., {Jungwiert}, B., {et~al.} 2007, \apj, 669, 801

\bibitem[{{Cappellari}(2008)}]{cappellari08}
{Cappellari}, M. 2008, \mnras, 390, 71

\bibitem[{{Cappellari} {et~al.}(2011){Cappellari}, {Emsellem}, {Krajnovi{\'c}},
  {McDermid}, {Scott}, {Verdoes Kleijn}, {Young}, {Alatalo}, {Bacon}, {Blitz},
  {Bois}, {Bournaud}, {Bureau}, {Davies}, {Davis}, {de Zeeuw}, {Duc},
  {Khochfar}, {Kuntschner}, {Lablanche}, {Morganti}, {Naab}, {Oosterloo},
  {Sarzi}, {Serra}, \& {Weijmans}}]{cappellari11a}
{Cappellari}, M., {Emsellem}, E., {Krajnovi{\'c}}, D., {et~al.} 2011, \mnras,
  413, 813

\bibitem[{{Cappellari} {et~al.}(2013){Cappellari}, {Scott}, {Alatalo}, {Blitz},
  {Bois}, {Bournaud}, {Bureau}, {Crocker}, {Davies}, {Davis}, {de Zeeuw},
  {Duc}, {Emsellem}, {Khochfar}, {Krajnovi{\'c}}, {Kuntschner}, {McDermid},
  {Morganti}, {Naab}, {Oosterloo}, {Sarzi}, {Serra}, {Weijmans}, \&
  {Young}}]{cappellari13a}
{Cappellari}, M., {Scott}, N., {Alatalo}, K., {et~al.} 2013, \mnras, 432, 1709

\bibitem[{{Chae} {et~al.}(2020){Chae}, {Lelli}, {Desmond}, {McGaugh}, {Li}, \&
  {Schombert}}]{chae20}
{Chae}, K.-H., {Lelli}, F., {Desmond}, H., {et~al.} 2020, \apj, 904, 51

\bibitem[{{Coleman} {et~al.}(2004){Coleman}, {Da Costa}, {Bland-Hawthorn},
  {Mart{\'{\i}}nez-Delgado}, {Freeman}, \& {Malin}}]{coleman04}
{Coleman}, M., {Da Costa}, G.~S., {Bland-Hawthorn}, J., {et~al.} 2004, \aj,
  127, 832

\bibitem[{{Cooper} {et~al.}(2011){Cooper}, {Mart{\'\i}nez-Delgado}, {Helly},
  {Frenk}, {Cole}, {Crawford}, {Zibetti}, {Carballo-Bello}, \&
  {GaBany}}]{cooper11}
{Cooper}, A.~P., {Mart{\'\i}nez-Delgado}, D., {Helly}, J., {et~al.} 2011,
  \apjl, 743, L21

\bibitem[{{Cullen} {et~al.}(2006){Cullen}, {Alexander}, \&
  {Clemens}}]{cullen06}
{Cullen}, H., {Alexander}, P., \& {Clemens}, M. 2006, \mnras, 366, 49

\bibitem[{{Deason} {et~al.}(2013){Deason}, {Van der Marel}, {Guhathakurta},
  {Sohn}, \& {Brown}}]{deason13}
{Deason}, A.~J., {Van der Marel}, R.~P., {Guhathakurta}, P., {Sohn}, S.~T., \&
  {Brown}, T.~M. 2013, \apj, 766, 24

\bibitem[{{Diemer} \& {Kravtsov}(2015)}]{diemer15}
{Diemer}, B. \& {Kravtsov}, A.~V. 2015, \apj, 799, 108

\bibitem[{{Dong-P{\'a}ez} {et~al.}(2021){Dong-P{\'a}ez}, {Vasiliev}, \&
  {Evans}}]{dp21}
{Dong-P{\'a}ez}, C.~A., {Vasiliev}, E., \& {Evans}, N.~W. 2021, \mnras
  [\eprint[arXiv]{2110.01060}]

\bibitem[{{Donlon} {et~al.}(2020){Donlon}, {Newberg}, {Sanderson}, \&
  {Widrow}}]{donlon20}
{Donlon}, Thomas, I., {Newberg}, H.~J., {Sanderson}, R., \& {Widrow}, L.~M.
  2020, \apj, 902, 119

\bibitem[{{Duc} {et~al.}(2015){Duc}, {Cuillandre}, {Karabal}, {Cappellari},
  {Alatalo}, {Blitz}, {Bournaud}, {Bureau}, {Crocker}, {Davies}, {Davis}, {de
  Zeeuw}, {Emsellem}, {Khochfar}, {Krajnovi{\'c}}, {Kuntschner}, {McDermid},
  {Michel-Dansac}, {Morganti}, {Naab}, {Oosterloo}, {Paudel}, {Sarzi}, {Scott},
  {Serra}, {Weijmans}, \& {Young}}]{duc15}
{Duc}, P.-A., {Cuillandre}, J.-C., {Karabal}, E., {et~al.} 2015, \mnras, 446,
  120

\bibitem[{{Dupraz} \& {Combes}(1986)}]{dc86}
{Dupraz}, C. \& {Combes}, F. 1986, \aap, 166, 53

\bibitem[{{Dupraz} \& {Combes}(1987)}]{dc87}
{Dupraz}, C. \& {Combes}, F. 1987, \aap, 185, L1

\bibitem[{{Ebrov{\'a}}(2013)}]{ebrovadiz}
{Ebrov{\'a}}, I. 2013, PhD thesis, Charles University Prague, arxiv:1312.1643

\bibitem[{{Ebrov{\'a}} {et~al.}(2020){Ebrov{\'a}}, {B{\'\i}lek},
  {Y{\i}ld{\i}z}, \& {Eli{\'a}{\v{s}}ek}}]{ebrova20}
{Ebrov{\'a}}, I., {B{\'\i}lek}, M., {Y{\i}ld{\i}z}, M.~K., \&
  {Eli{\'a}{\v{s}}ek}, J. 2020, \aap, 634, A73

\bibitem[{{Ebrov{\'a}} {et~al.}(2012){Ebrov{\'a}}, {J{\'\i}lkov{\'a}},
  {Jungwiert}, {K{\v{r}}{\'\i}{\v{z}}ek}, {B{\'\i}lek}, {Barto{\v{s}}kov{\'a}},
  {Skalick{\'a}}, \& {Stoklasov{\'a}}}]{ebrova12}
{Ebrov{\'a}}, I., {J{\'\i}lkov{\'a}}, L., {Jungwiert}, B., {et~al.} 2012, \aap,
  545, A33

\bibitem[{{Emsellem} {et~al.}(2011){Emsellem}, {Cappellari}, {Krajnovi{\'c}},
  {Alatalo}, {Blitz}, {Bois}, {Bournaud}, {Bureau}, {Davies}, {Davis}, {de
  Zeeuw}, {Khochfar}, {Kuntschner}, {Lablanche}, {McDermid}, {Morganti},
  {Naab}, {Oosterloo}, {Sarzi}, {Scott}, {Serra}, {van de Ven}, {Weijmans}, \&
  {Young}}]{emsellem11}
{Emsellem}, E., {Cappellari}, M., {Krajnovi{\'c}}, D., {et~al.} 2011, \mnras,
  414, 888

\bibitem[{{Famaey} {et~al.}(2018){Famaey}, {McGaugh}, \& {Milgrom}}]{famaey18}
{Famaey}, B., {McGaugh}, S., \& {Milgrom}, M. 2018, \mnras, 480, 473

\bibitem[{{Famaey} \& {McGaugh}(2012)}]{famaey12}
{Famaey}, B. \& {McGaugh}, S.~S. 2012, Living Reviews in Relativity, 15, 10

\bibitem[{{Fardal} {et~al.}(2008){Fardal}, {Babul}, {Guhathakurta}, {Gilbert},
  \& {Dodge}}]{fardal08}
{Fardal}, M.~A., {Babul}, A., {Guhathakurta}, P., {Gilbert}, K.~M., \& {Dodge},
  C. 2008, \apjl, 682, L33

\bibitem[{{Fardal} {et~al.}(2007){Fardal}, {Guhathakurta}, {Babul}, \&
  {McConnachie}}]{fardal07}
{Fardal}, M.~A., {Guhathakurta}, P., {Babul}, A., \& {McConnachie}, A.~W. 2007,
  \mnras, 380, 15

\bibitem[{{Fardal} {et~al.}(2012){Fardal}, {Guhathakurta}, {Gilbert},
  {Tollerud}, {Kalirai}, {Tanaka}, {Beaton}, {Chiba}, {Komiyama}, \&
  {Iye}}]{fardal12}
{Fardal}, M.~A., {Guhathakurta}, P., {Gilbert}, K.~M., {et~al.} 2012, \mnras,
  3140

\bibitem[{{Fensch} {et~al.}(2020){Fensch}, {Duc}, {Lim}, {Emsellem},
  {B{\'\i}lek}, {Durrell}, {Liu}, {Peng}, \& {Smith}}]{fensch20}
{Fensch}, J., {Duc}, P.-A., {Lim}, S., {et~al.} 2020, \aap, 644, A164

\bibitem[{{Forbes} \& {Thomson}(1992)}]{ft92}
{Forbes}, D.~A. \& {Thomson}, R.~C. 1992, \mnras, 254, 723

\bibitem[{{Forbes} {et~al.}(1994){Forbes}, {Thomson}, {Groom}, \&
  {Williger}}]{forbes94}
{Forbes}, D.~A., {Thomson}, R.~C., {Groom}, W., \& {Williger}, G.~M. 1994, \aj,
  107, 1713

\bibitem[{{Foster} {et~al.}(2014){Foster}, {Lux}, {Romanowsky},
  {Mart{\'\i}nez-Delgado}, {Zibetti}, {Arnold}, {Brodie}, {Ciardullo},
  {GaBany}, {Merrifield}, {Singh}, \& {Strader}}]{foster14}
{Foster}, C., {Lux}, H., {Romanowsky}, A.~J., {et~al.} 2014, \mnras, 442, 3544

\bibitem[{{Gentile} {et~al.}(2011){Gentile}, {Famaey}, \& {de
  Blok}}]{gentile11}
{Gentile}, G., {Famaey}, B., \& {de Blok}, W.~J.~G. 2011, \aap, 527, A76

\bibitem[{{Helmi} {et~al.}(2003){Helmi}, {Navarro}, {Meza}, {Steinmetz}, \&
  {Eke}}]{helmi03}
{Helmi}, A., {Navarro}, J.~F., {Meza}, A., {Steinmetz}, M., \& {Eke}, V.~R.
  2003, \apjl, 592, L25

\bibitem[{{Hendel} \& {Johnston}(2015)}]{hendel15}
{Hendel}, D. \& {Johnston}, K.~V. 2015, \mnras, 454, 2472

\bibitem[{{Hernquist} \& {Quinn}(1987{\natexlab{a}})}]{hq87b}
{Hernquist}, L. \& {Quinn}, P.~J. 1987{\natexlab{a}}, \apj, 312, 17

\bibitem[{{Hernquist} \& {Quinn}(1987{\natexlab{b}})}]{hq87a}
{Hernquist}, L. \& {Quinn}, P.~J. 1987{\natexlab{b}}, \apj, 312, 1

\bibitem[{{Hernquist} \& {Quinn}(1988)}]{hq88}
{Hernquist}, L. \& {Quinn}, P.~J. 1988, \apj, 331, 682

\bibitem[{{Hernquist} \& {Weil}(1992)}]{hernquist92}
{Hernquist}, L. \& {Weil}, M.~L. 1992, \nat, 358, 734

\bibitem[{{Husemann} {et~al.}(2010){Husemann}, {S{\'a}nchez}, {Wisotzki},
  {Jahnke}, {Kupko}, {Nugroho}, \& {Schramm}}]{hus10}
{Husemann}, B., {S{\'a}nchez}, S.~F., {Wisotzki}, L., {et~al.} 2010, \aap, 519,
  A115

\bibitem[{{Kado-Fong} {et~al.}(2018){Kado-Fong}, {Greene}, {Hendel},
  {Price-Whelan}, {Greco}, {Goulding}, {Huang}, {Johnston}, {Komiyama}, \&
  {Lee}}]{kadofong18}
{Kado-Fong}, E., {Greene}, J.~E., {Hendel}, D., {et~al.} 2018, \apj, 866, 103

\bibitem[{{Krajnovi{\'c}} {et~al.}(2011){Krajnovi{\'c}}, {Emsellem},
  {Cappellari}, {Alatalo}, {Blitz}, {Bois}, {Bournaud}, {Bureau}, {Davies},
  {Davis}, {de Zeeuw}, {Khochfar}, {Kuntschner}, {Lablanche}, {McDermid},
  {Morganti}, {Naab}, {Oosterloo}, {Sarzi}, {Scott}, {Serra}, {Weijmans}, \&
  {Young}}]{krajnovic11}
{Krajnovi{\'c}}, D., {Emsellem}, E., {Cappellari}, M., {et~al.} 2011, \mnras,
  414, 2923

\bibitem[{{Lelli} {et~al.}(2016){Lelli}, {McGaugh}, \& {Schombert}}]{lelli16}
{Lelli}, F., {McGaugh}, S.~S., \& {Schombert}, J.~M. 2016, \aj, 152, 157

\bibitem[{{Lelli} {et~al.}(2017){Lelli}, {McGaugh}, {Schombert}, \&
  {Pawlowski}}]{lelli17}
{Lelli}, F., {McGaugh}, S.~S., {Schombert}, J.~M., \& {Pawlowski}, M.~S. 2017,
  \apj, 836, 152

\bibitem[{{Lim} {et~al.}(2017){Lim}, {Peng}, {Duc}, {Fensch}, {Durrell},
  {Harris}, {Cuillandre}, {Gwyn}, {Lan{\c{c}}on}, \&
  {S{\'a}nchez-Janssen}}]{lim17}
{Lim}, S., {Peng}, E.~W., {Duc}, P.-A., {et~al.} 2017, \apj, 835, 123

\bibitem[{{Lima Neto} {et~al.}(1999){Lima Neto}, {Gerbal}, \&
  {M{\'a}rquez}}]{limaneto99}
{Lima Neto}, G.~B., {Gerbal}, D., \& {M{\'a}rquez}, I. 1999, \mnras, 309, 481

\bibitem[{{Longobardi} {et~al.}(2015){Longobardi}, {Arnaboldi}, {Gerhard}, \&
  {Mihos}}]{lon15b}
{Longobardi}, A., {Arnaboldi}, M., {Gerhard}, O., \& {Mihos}, J.~C. 2015, \aap,
  579, L3

\bibitem[{{L{\"u}ghausen} {et~al.}(2015){L{\"u}ghausen}, {Famaey}, \&
  {Kroupa}}]{por}
{L{\"u}ghausen}, F., {Famaey}, B., \& {Kroupa}, P. 2015, Canadian Journal of
  Physics, 93, 232

\bibitem[{{Malin} \& {Carter}(1983)}]{MC83}
{Malin}, D.~F. \& {Carter}, D. 1983, \apj, 274, 534

\bibitem[{{Mancillas} {et~al.}(2019){Mancillas}, {Combes}, \&
  {Duc}}]{mancillas19}
{Mancillas}, B., {Combes}, F., \& {Duc}, P.~A. 2019, \aap, 630, A112

\bibitem[{{M{\'a}rquez} {et~al.}(2000){M{\'a}rquez}, {Lima Neto}, {Capelato},
  {Durret}, \& {Gerbal}}]{marquez00}
{M{\'a}rquez}, I., {Lima Neto}, G.~B., {Capelato}, H., {Durret}, F., \&
  {Gerbal}, D. 2000, \aap, 353, 873

\bibitem[{{Mart{\'\i}nez-Delgado} {et~al.}(2010){Mart{\'\i}nez-Delgado},
  {Gabany}, {Crawford}, {Zibetti}, {Majewski}, {Rix}, {Fliri},
  {Carballo-Bello}, {Bardalez-Gagliuffi}, {Pe{\~n}arrubia}, {Chonis}, {Madore},
  {Trujillo}, {Schirmer}, \& {McDavid}}]{md10}
{Mart{\'\i}nez-Delgado}, D., {Gabany}, R.~J., {Crawford}, K., {et~al.} 2010,
  \aj, 140, 962

\bibitem[{{McGaugh} \& {Milgrom}(2013)}]{anddwarfii}
{McGaugh}, S. \& {Milgrom}, M. 2013, \apj, 775, 139

\bibitem[{{McGaugh}(2016)}]{mcgaughcra}
{McGaugh}, S.~S. 2016, \apjl, 832, L8

\bibitem[{{McGaugh} {et~al.}(2016){McGaugh}, {Lelli}, \&
  {Schombert}}]{mcgaugh16}
{McGaugh}, S.~S., {Lelli}, F., \& {Schombert}, J.~M. 2016, Physical Review
  Letters, 117, 201101

\bibitem[{{Meidt} {et~al.}(2014){Meidt}, {Schinnerer}, {van de Ven},
  {Zaritsky}, {Peletier}, {Knapen}, {Sheth}, {Regan}, {Querejeta},
  {Mu{\~n}oz-Mateos}, {Kim}, {Hinz}, {Gil de Paz}, {Athanassoula}, {Bosma},
  {Buta}, {Cisternas}, {Ho}, {Holwerda}, {Skibba}, {Laurikainen}, {Salo},
  {Gadotti}, {Laine}, {Erroz-Ferrer}, {Comer{\'o}n}, {Men{\'e}ndez-Delmestre},
  {Seibert}, \& {Mizusawa}}]{meidt14}
{Meidt}, S.~E., {Schinnerer}, E., {van de Ven}, G., {et~al.} 2014, \apj, 788,
  144

\bibitem[{{Merrifield} \& {Kuijken}(1998)}]{mk98}
{Merrifield}, M.~R. \& {Kuijken}, K. 1998, \mnras, 297, 1292

\bibitem[{{Miki} \& {Umemura}(2018)}]{magi}
{Miki}, Y. \& {Umemura}, M. 2018, \mnras, 475, 2269

\bibitem[{{Milgrom}(1983)}]{milg83a}
{Milgrom}, M. 1983, \apj, 270, 365

\bibitem[{{Milgrom}(2010)}]{qumond}
{Milgrom}, M. 2010, \mnras, 403, 886

\bibitem[{{Milgrom}(2012)}]{milg12}
{Milgrom}, M. 2012, Physical Review Letters, 109, 131101

\bibitem[{{Milgrom}(2013)}]{milg13}
{Milgrom}, M. 2013, \prl, 111, 041105

\bibitem[{{Miller} {et~al.}(2017){Miller}, {Ahumada}, {Puzia}, {Candlish},
  {McGaugh}, {Mihos}, {Sanderson}, {Schirmer}, {Smith}, \& {Taylor}}]{mil17}
{Miller}, B., {Ahumada}, T., {Puzia}, T., {et~al.} 2017, Galaxies, 5, 29

\bibitem[{{Mo} {et~al.}(2010){Mo}, {van den Bosch}, \& {White}}]{mo10}
{Mo}, H., {van den Bosch}, F.~C., \& {White}, S. 2010, {Galaxy Formation and
  Evolution}

\bibitem[{{Nagesh} {et~al.}(2021){Nagesh}, {Banik}, {Thies}, {Kroupa},
  {Famaey}, {Wittenburg}, {Parziale}, \& {Haslbauer}}]{nagesh21}
{Nagesh}, S.~T., {Banik}, I., {Thies}, I., {et~al.} 2021, Canadian Journal of
  Physics, 99, 607

\bibitem[{{Navarro} {et~al.}(1996){Navarro}, {Frenk}, \& {White}}]{nfw}
{Navarro}, J.~F., {Frenk}, C.~S., \& {White}, S.~D.~M. 1996, \apj, 462, 563

\bibitem[{{Perret}(2016)}]{dice}
{Perret}, V. 2016, {DICE: Disk Initial Conditions Environment}

\bibitem[{{Pop} {et~al.}(2018){Pop}, {Pillepich}, {Amorisco}, \&
  {Hernquist}}]{pop18}
{Pop}, A.-R., {Pillepich}, A., {Amorisco}, N.~C., \& {Hernquist}, L. 2018,
  \mnras, 480, 1715

\bibitem[{{Prieur}(1988)}]{prieur88}
{Prieur}, J.-L. 1988, \apj, 326, 596

\bibitem[{{Prieur}(1990)}]{prieur90}
{Prieur}, J.~L. 1990, {Status of shell galaxies.}, ed. R.~{Wielen}, 72--83

\bibitem[{{Puzia} {et~al.}(2006){Puzia}, {Kissler-Patig}, \&
  {Goudfrooij}}]{puzia06}
{Puzia}, T.~H., {Kissler-Patig}, M., \& {Goudfrooij}, P. 2006, \apj, 648, 383

\bibitem[{{Quinn}(1983)}]{quinn83}
{Quinn}, P.~J. 1983, in Internal Kinematics and Dynamics of Galaxies, ed.
  E.~{Athanassoula}, Vol. 100, 347

\bibitem[{{Quinn}(1984)}]{quinn84}
{Quinn}, P.~J. 1984, \apj, 279, 596

\bibitem[{{Ramos Almeida} {et~al.}(2011){Ramos Almeida}, {Tadhunter}, {Inskip},
  {Morganti}, {Holt}, \& {Dicken}}]{ra11}
{Ramos Almeida}, C., {Tadhunter}, C.~N., {Inskip}, K.~J., {et~al.} 2011,
  \mnras, 410, 1550

\bibitem[{{Rampazzo} {et~al.}(2006){Rampazzo}, {Alexander}, {Carignan},
  {Clemens}, {Cullen}, {Garrido}, {Marcelin}, {Sheth}, \&
  {Trinchieri}}]{rampazzo06}
{Rampazzo}, R., {Alexander}, P., {Carignan}, C., {et~al.} 2006, \mnras, 368,
  851

\bibitem[{{Reduzzi} {et~al.}(1996){Reduzzi}, {Longhetti}, \&
  {Rampazzo}}]{reduzzi96}
{Reduzzi}, L., {Longhetti}, M., \& {Rampazzo}, R. 1996, \mnras, 282, 149

\bibitem[{{Renaud} {et~al.}(2016){Renaud}, {Famaey}, \& {Kroupa}}]{renaud16}
{Renaud}, F., {Famaey}, B., \& {Kroupa}, P. 2016, \mnras, 463, 3637

\bibitem[{{Richtler} {et~al.}(2008){Richtler}, {Schuberth}, {Hilker}, {Dirsch},
  {Bassino}, \& {Romanowsky}}]{richtler08}
{Richtler}, T., {Schuberth}, Y., {Hilker}, M., {et~al.} 2008, \aap, 478, L23

\bibitem[{{Romanowsky} {et~al.}(2012){Romanowsky}, {Strader}, {Brodie},
  {Mihos}, {Spitler}, {Forbes}, {Foster}, \& {Arnold}}]{romanowsky12}
{Romanowsky}, A.~J., {Strader}, J., {Brodie}, J.~P., {et~al.} 2012, \apj, 748,
  29

\bibitem[{{Rong} {et~al.}(2018){Rong}, {Li}, {Wang}, {Gao}, {Li}, {Ge}, {Jing},
  {Pan}, {Fern{\'a}ndez-Trincado}, {Valenzuela}, \& {Ort{\'{\i}}z}}]{rong18}
{Rong}, Y., {Li}, H., {Wang}, J., {et~al.} 2018, \mnras, 477, 230

\bibitem[{{Salo} {et~al.}(2015){Salo}, {Laurikainen}, {Laine}, {Comer{\'o}n},
  {Gadotti}, {Buta}, {Sheth}, {Zaritsky}, {Ho}, {Knapen}, {Athanassoula},
  {Bosma}, {Laine}, {Cisternas}, {Kim}, {Mu{\~n}oz-Mateos}, {Regan}, {Hinz},
  {Gil de Paz}, {Menendez-Delmestre}, {Mizusawa}, {Erroz-Ferrer}, {Meidt}, \&
  {Querejeta}}]{salo15}
{Salo}, H., {Laurikainen}, E., {Laine}, J., {et~al.} 2015, \apjs, 219, 4

\bibitem[{{Samurovi{\'c}}(2014)}]{samur14}
{Samurovi{\'c}}, S. 2014, \aap, 570, A132

\bibitem[{{S{\'a}nchez} {et~al.}(2019){S{\'a}nchez}, {Avila-Reese},
  {Rodr{\'\i}guez-Puebla}, {Ibarra-Medel}, {Calette}, {Bershady},
  {Hern{\'a}ndez-Toledo}, {Pan}, \& {Bizyaev}}]{sanchez19}
{S{\'a}nchez}, S.~F., {Avila-Reese}, V., {Rodr{\'\i}guez-Puebla}, A., {et~al.}
  2019, \mnras, 482, 1557

\bibitem[{{Sanderson} \& {Helmi}(2013)}]{sh13}
{Sanderson}, R.~E. \& {Helmi}, A. 2013, \mnras, 435, 378

\bibitem[{{Schiminovich} {et~al.}(1997){Schiminovich}, {van Gorkom}, {van der
  Hulst}, {Oosterloo}, \& {Wilkinson}}]{schiminovich97}
{Schiminovich}, D., {van Gorkom}, J., {van der Hulst}, T., {Oosterloo}, T., \&
  {Wilkinson}, A. 1997, in Astronomical Society of the Pacific Conference
  Series, Vol. 116, The Nature of Elliptical Galaxies; 2nd Stromlo Symposium,
  ed. M.~{Arnaboldi}, G.~S. {Da Costa}, \& P.~{Saha}, 362

\bibitem[{{Schweizer} \& {Seitzer}(1988)}]{ss88}
{Schweizer}, F. \& {Seitzer}, P. 1988, \apj, 328, 88

\bibitem[{{Seguin} \& {Dupraz}(1996)}]{segdup96}
{Seguin}, P. \& {Dupraz}, C. 1996, \aap, 310, 757

\bibitem[{{Shelest} \& {Lelli}(2020)}]{shelest20}
{Shelest}, A. \& {Lelli}, F. 2020, \aap, 641, A31

\bibitem[{{Sheth} {et~al.}(2010){Sheth}, {Regan}, {Hinz}, {Gil de Paz},
  {Men{\'e}ndez-Delmestre}, {Mu{\~n}oz-Mateos}, {Seibert}, {Kim},
  {Laurikainen}, {Salo}, {Gadotti}, {Laine}, {Mizusawa}, {Armus},
  {Athanassoula}, {Bosma}, {Buta}, {Capak}, {Jarrett}, {Elmegreen},
  {Elmegreen}, {Knapen}, {Koda}, {Helou}, {Ho}, {Madore}, {Masters},
  {Mobasher}, {Ogle}, {Peng}, {Schinnerer}, {Surace}, {Zaritsky},
  {Comer{\'o}n}, {de Swardt}, {Meidt}, {Kasliwal}, \& {Aravena}}]{sheth10}
{Sheth}, K., {Regan}, M., {Hinz}, J.~L., {et~al.} 2010, \pasp, 122, 1397

\bibitem[{{Sikkema} {et~al.}(2007){Sikkema}, {Carter}, {Peletier}, {Balcells},
  {Del Burgo}, \& {Valentijn}}]{sikkema07}
{Sikkema}, G., {Carter}, D., {Peletier}, R.~F., {et~al.} 2007, \aap, 467, 1011

\bibitem[{{Tal} {et~al.}(2009){Tal}, {van Dokkum}, {Nelan}, \&
  {Bezanson}}]{tal09}
{Tal}, T., {van Dokkum}, P.~G., {Nelan}, J., \& {Bezanson}, R. 2009, \aj, 138,
  1417

\bibitem[{{Thronson} {et~al.}(1989){Thronson}, {Bally}, \& {Hacking}}]{thr89}
{Thronson}, Harley~A., J., {Bally}, J., \& {Hacking}, P. 1989, \aj, 97, 363

\bibitem[{{Tian} {et~al.}(2020){Tian}, {Umetsu}, {Ko}, {Donahue}, \&
  {Chiu}}]{tian20}
{Tian}, Y., {Umetsu}, K., {Ko}, C.-M., {Donahue}, M., \& {Chiu}, I.~N. 2020,
  \apj, 896, 70

\bibitem[{{Tiret} \& {Combes}(2008)}]{tiret08}
{Tiret}, O. \& {Combes}, F. 2008, in Astronomical Society of the Pacific
  Conference Series, Vol. 396, Formation and Evolution of Galaxy Disks, ed.
  J.~G. {Funes} \& E.~M. {Corsini}, 259

\bibitem[{{Turnbull} {et~al.}(1999){Turnbull}, {Bridges}, \&
  {Carter}}]{turnbull99}
{Turnbull}, A.~J., {Bridges}, T.~J., \& {Carter}, D. 1999, \mnras, 307, 967

\bibitem[{{Vakili} {et~al.}(2017){Vakili}, {Kroupa}, \& {Rahvar}}]{vakili17}
{Vakili}, H., {Kroupa}, P., \& {Rahvar}, S. 2017, \apj, 848, 55

\bibitem[{{Weil} \& {Hernquist}(1993)}]{weil93}
{Weil}, M.~L. \& {Hernquist}, L. 1993, \apj, 405, 142

\bibitem[{{Wilkinson} {et~al.}(1987){Wilkinson}, {Sparks}, {Carter}, \&
  {Malin}}]{wilkinson87}
{Wilkinson}, A., {Sparks}, W.~B., {Carter}, D., \& {Malin}, D.~A. 1987, in
  Structure and Dynamics of Elliptical Galaxies, ed. P.~T. {de Zeeuw}, Vol.
  127, 465

\bibitem[{{Young} {et~al.}(2020){Young}, {Krajnovi{\'c}}, {Duc}, \&
  {Serra}}]{young20}
{Young}, L.~M., {Krajnovi{\'c}}, D., {Duc}, P.-A., \& {Serra}, P. 2020, \mnras,
  495, 1433

\bibitem[{{Zhao}(2005)}]{zhao05}
{Zhao}, H.~S. 2005, \aap, 444, L25

\end{thebibliography}


\appendix

\section{Details of the simulations}
\label{app:details}
{ The self-consistent simulations were performed using the adaptive-mesh-refinement code Phantom of RAMSES \citep{por}. It can work both with Newtonian and MOND gravity.  The numerical parameters of the code were the same for all simulations we made, both Newtonian and MOND. They are listed in \tab{code}.}

The Cartesian coordinate system of the simulation was defined in the following way. The origin of the coordinate system coincided with the barycenter of the two galaxies. The $x$-axis coincided with the line connecting the centers of the merger progenitors at the beginning of the simulation and pointed toward the secondary. The $z$-axis pointed toward the observer and the $y$-axis was chosen to obtain a right-handed coordinate system.

\begin{table}
\caption{Setup of the PoR code used in all Newtonian and MOND simulations.}
\label{tab:code}
\begin{tabular}{l|l}
\hline\hline
Property/parameter  & Value                \\
\hline
\texttt{levelmin} & 7\\
\texttt{levelmax} & 14 \\
\texttt{boxlen} & 2\,Mpc \\
Maximum resolution & 0.12\,kpc \\
\texttt{mass\_sph}  & 1 \\
\texttt{m\_refine} & $200\,M_\sun$ \\
\hline
\end{tabular}
\end{table}

\begin{table}
\caption{Initial conditions in the MOND simulation}
    \label{tab:MONDsim}
    \begin{tabular}{lll}
    \hline
    \multicolumn{3}{c}{Stars} \\\hline
     & Primary & Secondary \\\hline\hline
    Profile & S\'ersic sphere & \makecell[l]{Approximate  \\ \hspace{0.5em} exponential \\ \hspace{0.5em} disk} \\
    $\log_{10}(\textrm{Mass}/M_\sun)$ &  10.521 & 9.82 \\
    Scale length [kpc] & 3.0 & 4.0 \\
    S\'ersic index & 2.0 & -- \\
    \makecell[l]{Sech$^2$-scale \\  \hspace{0.5em} height {[kpc]}}  & -- & 1.3\\
    Number of particles & $5\times 10^6$ & $5\times 10^6$\\
    \hline
\multicolumn{3}{c}{Orbital parameters} \\\hline\hline 
\makecell[l]{(position of secondary) $-$ \\ \hspace{0.5em}
(position of primary)} & \multicolumn{2}{l}{$(50,0,0)$\,kpc}\\
\makecell[l]{(velocity of secondary)$-$ \\ \hspace{0.5em}
(velocity of primary)} & \multicolumn{2}{l}{$(0,-5.0,0)$\,km\,s$^{-1}$} \\
Spin of secondary & \multicolumn{2}{l}{$(0,0,1)$} \\
\hline
\end{tabular}
\end{table}

\subsection{MOND simulations}
We started our MOND simulations by searching for the suitable initial conditions of the merger orbit using restricted three-body simulations (see, e.g., \citealp{ebrova12}). They are much faster than the self-consistent simulations. We inspected every simulation always at the time $T_\mathrm{1G}$ after the merger. These simulations suggested that we need a nearly radial orbit with galaxies starting with zero radial velocity in the distance of the brightest observed shell, that is, 30\,kpc, and that the disk of the secondary galaxy has to rotate around an axis parallel to the line of sight.

Then we proceeded to the self-consistent simulations.{  In general, the field of simulations of galaxies in MOND is not as advanced as with Newtonian gravity. There are currently no software tools to prepare stable initial models of galaxies with prescribed density profiles. It is currently necessary to improvise.} The initial models of the galaxies{ in this paper} were prepared in following way in the published MOND simulation. The density of the secondary was initiated as an exponential disk with a scale-length of 0.5\,kpc and zero thickness. The particles were assigned a systemic rotation velocity such that the centrifugal force balances the MOND acceleration. This was obtained using \equ{mond} from the Newtonian gravitational acceleration generated by the given exponential disk \citep{bt08}. In addition, we added to the velocities of particles a random component selected from an isotropic Gaussian distribution, whose dispersion was equal to 0.3 of the local rotation velocity. 
The disk was evolved in isolation for 2\,Gyr before being entered in simulation of the merger.{ This time period was determined by watching the morphology of the galaxy. We considered the galaxy to be virialized enough when its initial period of large morphological transformations ended.} The galaxy settled as a bulged exponential disk with a scale length of 4.0\,kpc, half-mass radius of 2.3\,kpc, and a scale height of the  $\sech^2$-profile  of 1.3\,kpc at the distance of one scale length from the center. These{ values follow}  the observed scaling relations  \citep{bizyaev02,lelli16}. The density of the elliptical galaxy was set up as a S\'ersic sphere with an effective radius of 3\,kpc and index of two. The particles were given random velocities from an isotropic Gaussian distribution with a velocity dispersion obtained by solving the Jeans equation. The primary, as well as the  secondary, consisted of $5\times 10^{6}$ particles. {The parameters of the initial galaxy models for the MOND simulation of collision are summarized in \tab{MONDsim}.}

The self-consistent simulations confirmed that we need a nearly radial collision, otherwise only a low number of well-defined shells appears. The characteristic combination of a narrow stream and shells observed in NGC\,474 formed only in collisions along radial or mildly retrograde orbits. In simulations with prograde orbits, the stream formed too wide. We explored the range of initial tangential velocities between about $\pm20$\,km\,s$^{-1}$ and separations of 40 and 50\,kpc. The simulations launched from 40\,kpc tended to produce an indistinct stream.  We also tried secondaries prepared in other ways, including the one described in \citet{nagesh21}, which makes use of an adjusted version of the DICE code \citep{dice}.
We found that at fixed orbital parameters of the encounter, the morphology of the shells is sensitive to the{ initial} structure of the secondary.

In the final simulation, the galaxies started 50\,kpc apart. The difference of the $y$-components of the velocities of the galaxies was 5.0\,km\,s$^{-1}$. The components of the velocities of the galaxies in the other directions were zero. The spin vector of the secondary pointed in the direction of the $z$-axis.{ The orbital initial conditions for the simulation are summarized in  \tab{MONDsim}.}

\subsection{Newtonian simulations}
{ The initial conditions of the Newtonian simulations are summarized in  \tab{dmsim}.} The stellar masses of the galaxies were the same as in the MOND simulation. The dark matter halos were modeled by NFW halos truncated at 100\,kpc.  The initial models were generated by the MAGI code \citep{magi} and were  evolved in isolation for 2\,Gyr before they were entered in the simulation of the collision. The scale radii and characteristic densities of the halos were{ derived as following the mean} dark matter scaling relations \citep{behroozi13,diemer15}. { We explored various initial distances of the primary and secondary. We obtained shells that are comparable in size with those observed for as large initial separation as possible. We avoided initial distances that are larger than the truncation radius of the halos because such simulations would not be able to treat dynamical friction correctly. The initial tangential velocity was left the same as for the final MOND simulation.}

{ We also performed simulations (which we do not publish here)  with dark halos chosen to mimic the MOND gravitational fields of the primary and secondary. We could not produce a tidal features resembling those in NGC\,474 for any orbital parameters. The shells were axially symmetric with a narrow azimuthal extent, rather resembling those in NGC\,7600 \citep[see, e.g.,][]{turnbull99,cooper11} or NGC\,3923 \citep[e.g.,][]{MC83,bil16}, even if the two galaxies started on a circular orbit 100\,kpc form each other. In addition, there was no good counterpart of the thin northern stream; the mergers rather produced wide tails.  The stellar components were the modeled just as in the published simulation. The halo of the primary was represented by a Burkert halo \citep{burkert95} with a core radius of 6.91\,kpc and characteristic density of $1.4\times 10^8\,M_\sun\,$kpc$^{-3}$. The halo of the secondary had a Burkert profile as well, with a core radius of 11.35\,kpc and a characteristic density of $1.2\times 10^8\,M_\sun\,$kpc$^{-3}$. Both halos were truncated at 100\,kpc. Dark matter particles had masses $5\times 10^5\,M_\sun$ and stellar particles $1\times 10^5\,M_\sun$ , and there were $1\times 10^7$ particles in total. }

\begin{table}
\caption{Initial conditions in the Newtonian simulation}
    \label{tab:dmsim}
    \begin{tabular}{lll}
    \hline
    \multicolumn{3}{c}{Stars} \\\hline
     & Primary & Secondary \\\hline\hline
    Profile & Hernquist sphere & \makecell[l]{Exponential \\ \hspace{0.5em} disk} \\
    $\log_{10}(\textrm{Mass}/M_\sun)$ &  10.521 & 9.82 \\
    Scale length [kpc] & 2 & 2 \\
    \makecell[l]{Exponential scale \\  \hspace{0.5em} height {[kpc]}}  & -- & 0.6\\
    Number of particles & 1,400,597 & 278,813\\
    \hline
    \multicolumn{3}{c}{DM halos} \\\hline
    & Primary & Secondary \\\hline\hline
    Profile & NFW & NFW \\
    \makecell[l]{Characteristic density \\  \hspace{0.5em}  {[$M_\sun\,$kpc$^{-3}$]}}
      & $2.76\times10^6$ & $3.87\times10^6$ \\
    Scale radius [kpc] & 31.1 & 17.7 \\
    Truncation radius [kpc] & 100 & 100 \\
    $\log_{10}(\textrm{Mass}/M_\sun)$ & 11.85 & 11.45 \\
    Number of particles & 5,938,956 & 2,381,632 \\\hline
\multicolumn{3}{c}{Orbital parameters} \\\hline\hline 
 \makecell[l]{(position of secondary) $-$\\ \hspace{0.5em}
 (position of primary)} & \multicolumn{2}{l}{$(100,0,0)$\,kpc}\\
\makecell[l]{(velocity of secondary) $-$ \\ \hspace{0.5em}
(velocity of primary)} &  \multicolumn{2}{l}{$(0,-5.0,0)$\,km\,s$^{-1}$} \\
Spin of secondary & \multicolumn{2}{l}{$(0,0,1)$} \\
\hline
\end{tabular}
\end{table}

\section{Verification of the analytic model of the shell propagation}
\label{app:ansim}

\begin{figure*}
        \centering
        \includegraphics[width=17cm]{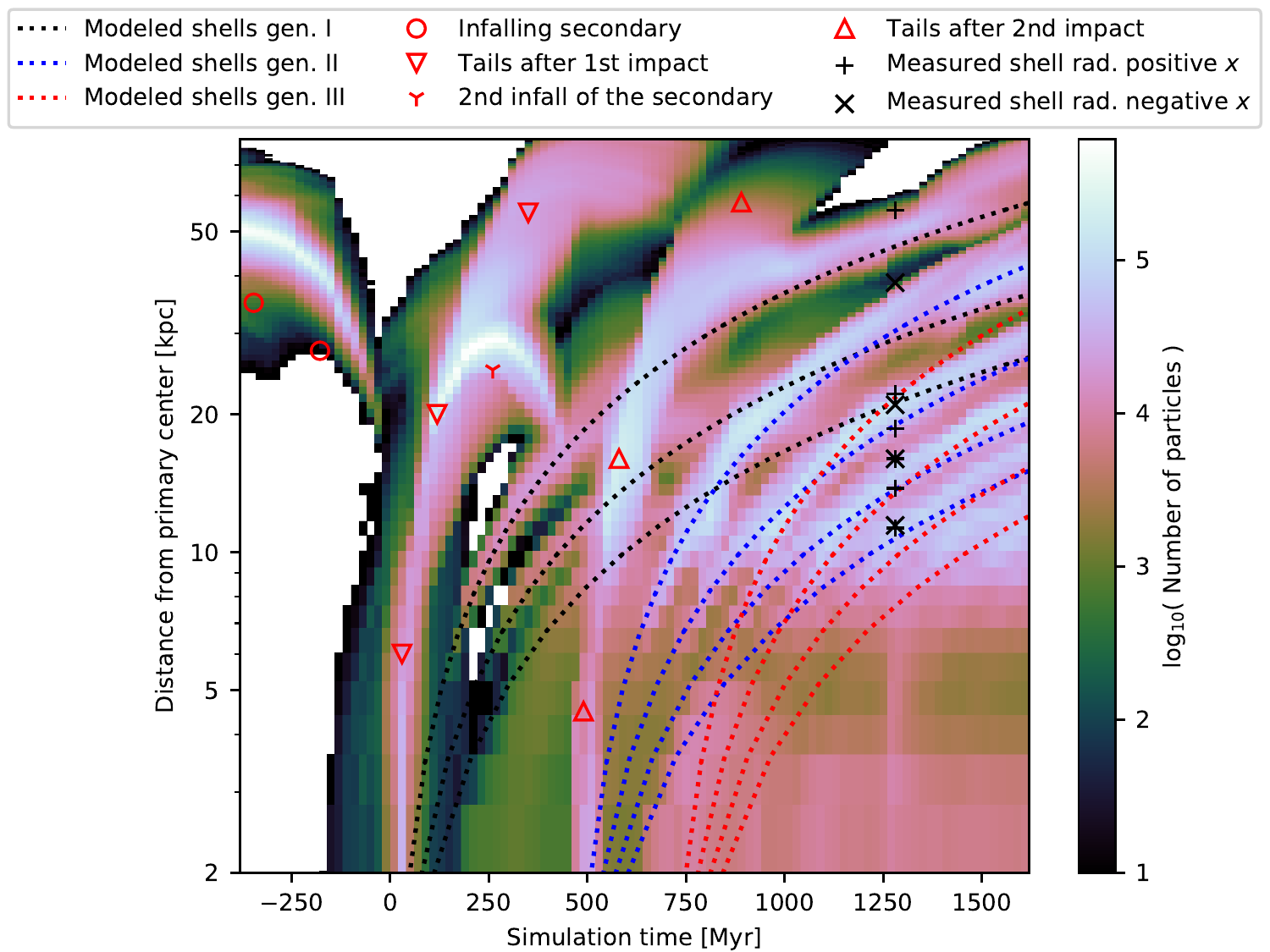}
        \caption{Comparison of the modeled evolution of shell radii with the MOND simulation. 
        }
        \label{fig:ansim}
\end{figure*}

The analytic models of the evolution of shell radii have not been verified by many self-consistent simulations. The only published comparison we are aware is \citet{ebrova20}.{ We made here another comparison  making use of the MOND simulation described in \sect{sim}. This is advantageous in the context of this paper because the simulated galaxy resembles the investigated real galaxy NGC\,474.}

{ To compare of the simulation with the analytic model, we} adopted the same procedure as in \citet{ebrova20}. In every time step of the simulation, we selected particles of the secondary galaxy that have their radial velocities with respect to the center of the primary galaxy close to zero.  These particles are those near the shell edges. The center of the primary galaxy was defined as its center of mass. For every time step we subsequently made a histogram of the selected particles according to their distance from the center of the primary galaxy. Combining the radial histograms for all time steps, we obtained{ the color map in } \fig{ansim}. 

The features seen in this plot correspond to shells, but also to other objects or structures. Near the start of the simulation, the structure marked by the circles corresponds to the infalling secondary. Its first pericenter occurs at the time of 0\,Myr. The structure marked by the down-pointing triangles corresponds to the surviving core of the secondary and to the tidal tails extending from it (see \fig{snap}). The feature marked by the three-point asterisk is the core of the secondary near its turnaround point. The second pericentric passage occurs 460\,Myr after the first. It is again followed by the formation of tails around the secondary, which corresponds to the feature marked by the upward-pointing triangles. 

Most of the remaining features in \fig{ansim} correspond to shells. The upper edges of the bands correspond to shell edges.  We manually measured the radii of the shells in an image of the simulated galaxy at the time \tig, in the same way as we did it for the real galaxy.  The image was very similar to that in the upper right  panel of \fig{morph}. The radii of the shells are marked in \fig{ansim} by the pluses for the shells lying on the positive part of the $x$-axis and by the $\text{}$crosses for those on the negative part.  

{ We then compared the measured radii with the analytic models of shell evolution to evaluate the precision of the models. We used the MOND gravitational potential from \sect{mondpot}. The models tell us the radii of the shell at a certain time after the pericenter of the galaxies when the particles in the shells were released from the secondary. There were several pericentric approaches of the galaxies in the simulation before the secondary dissolved. We thus plotted in \fig{ansim} the model of shell evolution three times, so that the time origins of the models coincide with the instants of the first three pericenters of the galaxies. The three shifted copies of the model of shell evolution correspond to the expected evolution of shell radii in the first three shell generations. The shells belonging to the same generation are distinguished by the different colors of the dotted curves in \fig{ansim}  The pericenters of the galaxies happened in the simulation times of 0, 460 and 700\,Myr.}

{ The first look reveals that the analytic model gives a good estimate of the positions of the shells because each of the branches in the simulated data is traced by some of the curves of the model. We made a more quantitative check{ for the shells whose} radii measured in the simulation at the time \tig.{ For these shells, we had to find the serial and generation numbers.} We did this first by matching each of the measured points to one of the bright bands in \fig{ansim} and then by matching of the bright bands to the analytic models of shell evolution. It was then possible to compare the analytic models  to the measurements.{ When comparing the measured shell radii to the models,} it was necessary to remember that some of the simulated shells that we could see in the image at the time \tig ~were probably blends of shells coming from different generations, as  \fig{ansim} suggests.{ Moreover, we} recall that some pairs of the simulated shells are probably single shells that encircle the galaxy because of the apsidal twist of the stars that make the shells up. These apparent pairs of shells can be recognized because the two shells lie on opposite sides of the galaxy and have nearly identical radii. After determining the identification of the simulated shells, we found that the analytic model agrees with the shell radii measured in the simulation  with a precision better than around 0.1\,dex.} The analytic models do not provide exact predictions because of the self-gravity of the accreted material and the nonzero size of the secondary before the encounter. The magnitude of the deviations likely depends on the internal structure of the secondary before the merger. 

The second largest shell marked by the cross is the counterpart of the observed shell $B$. It seems that the shell is a blend of the second shell of the first generation and the first shell of the second generation.{ This indicates that the observed shell $B$ could be such a blend as well.}

\begin{figure*}
        \centering
        \includegraphics[width=17cm]{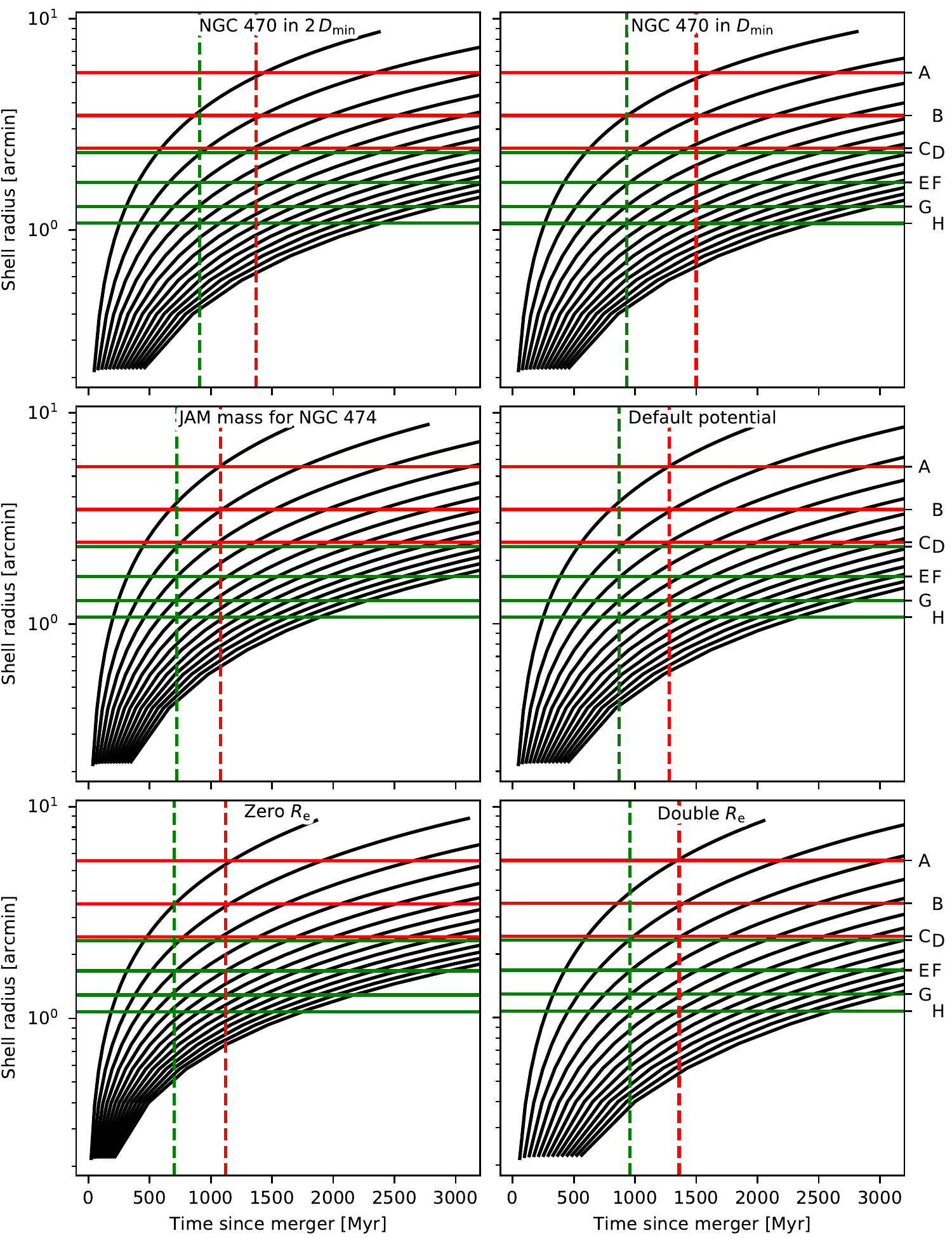}
        \caption{Comparison of  the observed shell radii with  the modeled evolution of shell radii for alternative{ MOND} gravitational potentials of NGC\,474. The meaning is analogous to \fig{shcomp}, but here we explored a larger variety of gravitational potentials. Top left:{ External field of NGC\,470 is taken into account}. The neighbor is assumed to be at a distance of $2D_\mathrm{min}$. Top right: The same, but for the distance of NGC\,470 of $D_\mathrm{min}$.{ Middle} left: MOND gravitational potential for no external field assuming mass derived from Jeans anisotropic modeling \citep{cappellari13a}.{ Middle right: Default MOND potential. Bottom left: Whole mass of the galaxy is considered to be concentrated in a single point to demonstrate the level of sensitivity of the shell evolution models with respect to the modeling of the mass distribution of the galaxy. Bottom right: Effective radius of the galaxy in the default model was multiplied by two for the same reason. }
        }
        \label{fig:shcomp-alt}
\end{figure*}

\begin{table}
\caption{Estimates of ages of shell generations in NGC\,474 for various MOND gravitational potentials.}            
\label{tab:MONDmod} 
\centering                                  
\begin{tabular}{llll}          
\hline\hline                       
Name & \tig & \tiig & Notes\\  
\hline                               
Default & 1280 & 870 & \makecell[l]{$\log_{10}\frac{M_*}{M_\sun}  = 10.6$,\\ 
\hspace{0.5em} $R_\mathrm{e}$ = 7.6\,kpc, \\
\hspace{0.5em} no NGC\,470} \\
NGC\,470 in $2\,D_\mathrm{min}$ & 1370 & 910 &  \\
NGC\,470 in $D_\mathrm{min}$ & 1500 & 930 &  \\
JAM mass for NGC 474 & 1080 & 720 & $\log_{10}\frac{M_*}{M_\sun}  = 10.87$ \\
Zero $R_\mathrm{e}$ & 1120 & 700 &  $R_\mathrm{e}$ = 0\,kpc \\
Double $R_\mathrm{e}$ & 1360 & 960 &  $R_\mathrm{e}$ = 15.3\,kpc \\
\hline                                            
\end{tabular}
\end{table}

\section{Alternative shell identification for  MOND?}
\label{app:alternative}
{ In \sect{ident} we presented for the MOND potential a shell identification that we found most probable. Here we discuss an alternative shell identification that is allowed by the shell identification method, but is disfavored by other constraints. In particular, the} modeled shell evolution would be consistent with the radii of all of the considered shells if they were formed in a single generation 870\,Myr ago, that is, at the time $T_\mathrm{2G}$ defined in \sect{ident}. The two shell pairs $C-D$ and $E-F$ have to be identified as two single shells with a surface brightness varying in azimuth. The brightest shell, $B$, gets the serial number one instead of two assigned in the most likely shell identification. 

Under such a scenario, the irregular structure $A$ is not a shell. This contradicts our { MOND} simulation, which shows that the stream is connected to shell 1,{ which thus corresponds to the observed structure $A$.} On the other hand, we did not explore the whole space of initial conditions. Also, this time of merger is not consistent with the theoretical considerations regarding the time delay between the merger and the central star formation. An actual hydrodynamical simulation is missing, however. In all our simulations, the shells were formed in more than one generation. This might again be a consequence of the small part of the parameter space we explored because the number of generations that form depends on the concentration of the secondary \citep{segdup96}.{ In principle, it would be} possible to distinguish observationally between the adopted and this alternative shell identification using high-resolution measurements of the shell spectra. The line-of-sight velocity distribution provides information about the shell expansion velocity, and from this, assuming the knowledge of the gravitational potential of the galaxy, the shell serial numbers can be derived \citep{ebrova12, bilcjp,bil15}.

\section{Can results change with alternative  MOND gravitational potentials?}
\label{app:other}
We based our interpretation of the formation of NGC\,474 on the assumption of the gravitational potential introduced in \sect{mondpot}.{  We call this the ``default'' hereafter. In this section}, we explore other choices of the potential.{ The ages of shell generations are listed in \tab{MONDmod} for all of the considered gravitational potentials in this section.} We begin with the exploration of the role of the external field effect on the MOND gravitational potential. This is important because NGC\,474 has a massive neighbor, NGC\,470. MOND is a nonlinear theory, meaning that the internal dynamics of a system is influenced by the presence of the external field, even if the latter is homogeneous. Observational signs of the external field effect have indeed been reported \citep{milg83a,anddwarfii,milg13,mcgaughcra,caldwell17,famaey18,chae20}. Here, we evaluate the external field effect with the 1D approximation \citep{famaey12},
\begin{equation}
     a = a_\mathrm{N}\nu\left(\frac{a_\mathrm{N}+a_\mathrm{N, ext}}{a_0}\right) + a_\mathrm{N,ext}\left[\nu\left(\frac{a_\mathrm{N}+a_\mathrm{N,ext}}{a_0}\right)-\nu\left(\frac{a_\mathrm{N,ext}}{a_0}\right)\right],
     \label{eq:efe}
\end{equation}
where $a_\mathrm{N,ext}$ denotes the magnitude of the external field calculated in the Newtonian way from the distribution of baryons. Using \equ{efe} instead of \equ{mond}, we could again model the evolution of shell radii in NGC\,474 taking the external field effect into account.

{ The largest regular shell of NGC\,474, $B$, shows no signs of tidal deformations. We thus used its size, that is, 31.3\,kpc, as the lowest acceptable tidal radius of NGC\,474.  The} MOND expression for the tidal radius \citep[Eq. (12)]{zhao05} implies a minimum separation of the galaxies of $D_\mathrm{min} = 95\,$kpc, that is, about twice the projected size. If the neighbor lies at a distance of $2D_\mathrm{min}$, the Newtonian external field amounts to $a_\mathrm{N, ext} = 2\times10^{-3}a_0$ and the tidal radius is 63\,kpc. This is still 1.3 times more than the radius of the irregular shell $A$. The comparison of the modeled and observed shell radii is shown in the top left panel of \fig{shcomp-alt}. The model for the { default MOND} gravitational field is shown again in the { middle right} panel of \fig{shcomp-alt}. This external field has virtually no influence on the estimated times since the two encounters with the secondary: the first occurred only 90\,Myr sooner, 1370\,Myr ago, and the second even only 40\,Myr sooner, that is,  910\,Myr ago. 

The external field reaches the maximum admissible value for the separation of NGC\,474 and NGC\,470 of $D_\mathrm{min}$. Then $a_\mathrm{N, ext} = 8\times10^{-3}a_0$ and the tidal radius has the diameter of shell $B$, that is, the irregular shell $A$ is already being stripped and should not be included in the comparisons with the shell propagation model because we neglected the tidal force in it. Even in this extreme case, the ages of the two shell generations are similar to the case without the external field. The fist generation is older by 220\,Myr (age of about 1500\,Myr) and the second is just 60\,Myr older (age of 930\,Myr). It is also possible to identify all shells as formed in one generation that is 930\,Myr old. This all demonstrates that the external field effect is not a substantial source of uncertainty for estimating the age of the shells of NGC\,474. 

We discuss an alternative stellar mass of the galaxy. A dynamical measurement of the $R$-band mass-to-light ratio of the galaxy, $(M/L)_e$ , was obtained from Jeans anisotropic modeling  \citep{cappellari08} measured by  \citet{cappellari13a}.{ The stellar mass can be obtained by multiplying  $(M/L)_e$ first by the total luminosity of the galaxy and then by the factor of 0.87 to account for the typical contribution of dark matter; see \citet{cappellari13a}. We obtained the stellar mass of the galaxy as $10^{10.87}\,M_\sun$, that is, 1.3 times} higher than the{ adopted mass that we used to derive the default MOND gravitational potential}. The adopted mass was chosen for several reasons. It was derived in the S4G survey, which was directly designed to measure the stellar masses of galaxies. It exploited two-band near-infrared photometry \citep{sheth10}. The mass-to-light ratio in the near-infrared {is not much sensitive} to the age of stellar population and dust obscuration \citep{meidt14}. The dynamical mass seems less reliable. The Jeans anisotropic modeling is based on the assumption of axial symmetry of the galaxy, but the photometric and kinematic axes in NGC\,474 are misaligned \citep{krajnovic11}.
In addition, the dynamical estimate is based on the observations of only a relatively small portion of the galaxy near its center. Namely, the radius of the probed region was only about 0.7 of the effective radius of the galaxy \citep{cappellari13a}.  Finally,  the estimate might have been distorted by the presence of the two populations with different ages in the center of the galaxy \citep{fensch20} that probably do not share their kinematics. 

Nevertheless, in the bottom left panel of \fig{shcomp-alt}, we explore how our estimates of the timing of the formation of the shells change if this alternative stellar mass is assumed. The results are very similar to those with the default gravitational potential, but the shell system is somewhat younger. The time since the formation of the{ first generation is 1080\,Myr (200\,Myr younger than for the default potential) and the time since the formation of the second is 720\,Myr (150\,Myr younger). The moment of the first impact coincides with the onset of the central star formation and the second impact with the second peak of central star formation (\fig{events}). The age of the cluster GC1 is still consistent with the time of the first impact, and the age of GC2 becomes closer to the second impact, but they are still more than $2\,\sigma$ inconsistent. While this alternative stellar mass still seems consistent with the ages of the young stellar populations in NGC\,474, we prefer the photometric mass for the reasons explained in the previous paragraph.}

{ Our default model of gravitational potential was based on a distribution of the stellar mass of the galaxy that was obtained by fitting the image of the galaxy by a S\'ersic sphere. Such fits never are absolutely precise. We therefore inspected the robustness of the estimates of the ages of the shell generations with respect to the mass distribution within the galaxy. We considered two additional models. In the first model, the effective radius of the galaxy was assumed to be zero, that is, we replaced the galaxy by a point mass. In the other model, we multiplied the observed effective radius (\sect{obs}) by two. The resulting models of shell evolution are compared to the observed shell radii in the two panels in the last row of \fig{shcomp-alt}. These radical changes of mass distribution have a rather mild impact on the estimates of the ages of the shell generations. With zero effective radius, the age of the first generation is 1120\,Myr (160\,Myr younger than with the default potential) and the age of the second is 700\,Myr (170\,Myr younger). For the double effective  radius, the age of the first shell generation is 1360\,Myr (80\,Myr older than with the default potential) and that of the second is 960\,Myr (80\,Myr older).} Finally, we note that none of the models of shell evolution considered in this paper supports the formation of the system more than 2\,Gyr ago, in contrast to how \citet{alabi20} interpreted the measured star formation history of shell $B$.

\end{document}